\documentclass[aps,prapplied,twocolumn,superscriptaddress]{revtex4-2}
\usepackage{graphicx}
\usepackage{amssymb,amsfonts,amsmath}
\usepackage{graphicx}
\usepackage{color}
\usepackage[usenames,dvipsnames]{xcolor}
\usepackage{epstopdf}
\usepackage[normalem]{ulem}
\usepackage{bm}
\usepackage{bbold}
\usepackage{float}
\usepackage{dsfont}
\usepackage{algorithm}
\usepackage{algpseudocode}

\renewcommand{\i}{{\rm i}}
\newcommand{\ket}[1]{|#1\rangle}
\newcommand{\bra}[1]{\langle #1|}

\newcommand{\norm}[1]{\left\lVert#1\right\rVert}
\usepackage[colorlinks=true,citecolor=Cerulean,linkcolor=RubineRed,urlcolor=Cerulean]{hyperref}
\hypersetup{breaklinks=true}
\begin{document}
\title{Software Framework for Optically Accessible Quantum Memories Using Group-IV Color Centers in Diamond}

\author{Yannick Strocka}
\affiliation{Department of Physics, Humboldt-Universität zu Berlin, 12489 Berlin, Germany}

\author{Mohamed Belhassen}
\affiliation{Department of Physics, Humboldt-Universität zu Berlin, 12489 Berlin, Germany}

\author{Tim Schr{\"o}der}
\affiliation{Department of Physics, Humboldt-Universität zu Berlin, 12489 Berlin, Germany}
\affiliation{Ferdinand-Braun-Institut, Leibniz-Institut für Höchstfrequenztechnik, 12489 Berlin, Germany}

\author{Gregor Pieplow*}
\affiliation{Department of Physics, Humboldt-Universität zu Berlin, 12489 Berlin, Germany}

\begin{abstract}
In the rapidly evolving field of quantum technology, the precise and detailed description of quantum components is not just a necessity but the foundation for advancing research, development, and applications. Optically accessible quantum memories are key building blocks for devices such as quantum repeaters and two-factor authentication. The memory we describe here is based on a Group-IV-vacancy color center (SiV$^-$/SnV$^-$) coupled to a highly efficient cavity. It leverages state-dependent reflection from the cavity and implements high-fidelity fractional single qubit gates via a train of optical $\pi/8$ pulses. We also describe its operation under microwave control, further extending our analysis. Our primary contribution in this work is the integration of this device model into a standardized software framework for quantum memory architectures.\\\\
\textbf{Key words:} quantum communication, quantum optics, quantum control  
\end{abstract}

\maketitle
\section{Introduction}
\begin{figure*}
    \centering
    \includegraphics[width=\linewidth]{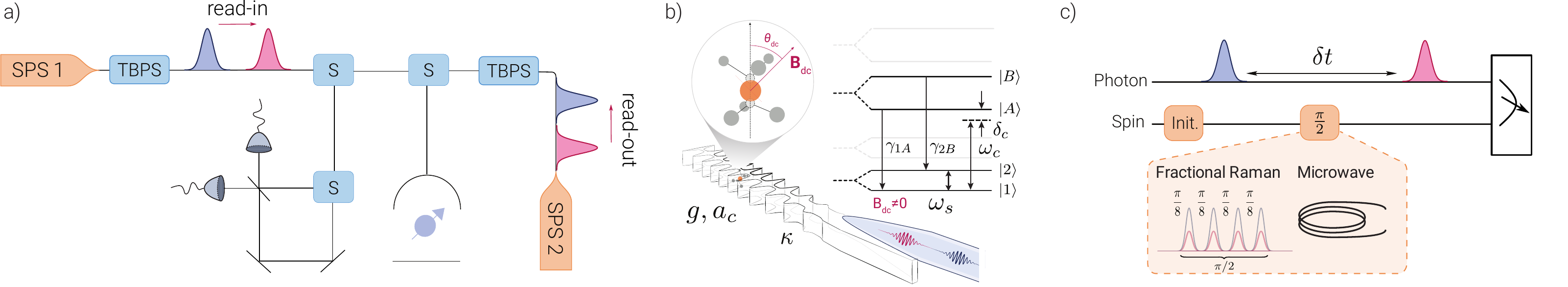}
    \caption{a) A time-bin qubit $\ket{\psi}$ is stored in a Sawfish \cite{bopp_sawfish_2024, pregnolato_fabrication_2024} spin-photon interface that hosts a G4V. Storage is achieved via the electronic spin. Fast switches (S) are deployed throughout the circuit to enable efficient measurement and routing of the photonic state. The $X$-basis measurement is performed using a fast switch integrated with an imbalanced Mach-Zehnder interferometer. The storage procedure is completed once the photonic qubit emitted from the single-photon source (SPS 1) is measured in the $X$-basis. For retrieval, the spin state is read out by entangling it with a photon from an additional single-photon source (SPS 2) after a time-bin preparation stage (TBPS) \cite{Lee2018, Bouchard2022, Yu2025}. A subsequent $Z$-basis measurement on the spin is performed. b) Using the SiV$^-/$SnV$^{-}$ center as a G4V, spin-photon entanglement is mediated through a Sawfish cavity-to-fiber interface \cite{bopp_sawfish_2024, pregnolato_fabrication_2024}. The SiV$^-$/SnV$^{-}$ is modeled as a four-level system characterized by its spontaneous emission rates $\gamma_{1A},\gamma_{2B}$, spectral contrast $\Delta\omega_s=\omega_{2B}-\omega_{1A}$ spin splitting $\omega_s$, and atomic transition frequencies $\omega_{1A},\omega_{2B}$, while interacting with a cavity mode $a_c$ at frequency $\omega_c$ for the spin-dependent reflection steps. c) 
    Entanglement is generated employing a reflection scheme containing spin-dependent reflection steps and a $\pi/2$ rotation where the spin is initialized in the state $\ket{1}$. The $\pi/2$ rotation is applied to the spin before the final reflection event. This rotation can be implemented either using a sequence of four $\pi/8$ optical Raman pulses or utilizing microwave control. Figures b) and c) are adopted from \cite{strocka_token_2025}.}
    \label{Figure:scheme}
\end{figure*}

Optically accessible quantum memories that can store and retrieve photonic qubits become of increasing interest in the era of quantum communication \cite{Gisin2007,Chen2021,Couteau2023}, computing \cite{Gill2021,Sood2024,Nimbe2021} and sensing \cite{Degen2017,Zhang2021,Chang2023}. Below, we highlight key applications: First generation quantum repeaters, a key component in scalable quantum networks, rely on a quantum memory to extend entanglement over long distances \cite{vanLoock2020,Yan2021,Azuma2023}. Beyond enabling long-distance entanglement, quantum memories also play a crucial role in quantum security applications \cite{Portmann2022} such as quantum authentication \cite{Dutta2022,strocka_token_2025}.
While quantum memories are important for quantum security and communication they are also highly relevant for scalable quantum computing \cite{Kyaw2015}, and especially for the development of a quantum random-access-memory (QRAM) \cite{qram2023}. Finally, the field of quantum sensing benefits from the use of quantum memories since they can help in increasing the sensor's sensitivity \cite{Zaiser2016}.

There are various types of quantum memories: The atomic-ensemble based quantum memories can map a photonic qubit onto a collective-spin excitation shared between two atomic ensembles \cite{Tanji2009}, or onto a spatially multiplexed cold atomic ensemble \cite{VernazGris2018} achieving storage times in the range of tens of microseconds. Another type of memory is based on solid-state systems: Prominent examples are rare-earth ion-doped crystals \cite{LagoRivera2021} or color centers in diamond \cite{Bradley2022, Bhaskar2020}, which are able to store quantum information using methods such as the atomic frequency comb (AFC) protocol \cite{LagoRivera2021,Bonarota2010} or by using a spin-photon interface and long lived nuclear spins in the diamond lattice  \cite{Bradley2022}. Superconducting quantum memories are also being actively investigated \cite{Wendin2017,Bao2021,Matanin2023,Ganjam2024}. Although superconducting qubits have not yet been used to store individual photonic qubits, neither directly nor via transduction \cite{Milul2023,Wang2022}, they have instead relied on superconducting waveguides for networking \cite{magnard_microwave_2020}. Recent progress has nevertheless been made in all-optical single-shot readout \cite{Arnold2025}. Finally, hybrid quantum memories combine physically distinct systems to push the scheme’s performance beyond that of the individual components. Take for example storage and retrieval of photons from a quantum dot single-photon source interfaced with an atomic quantum memory \cite{Thomas2024}, based on the Off-Resonant Cascaded Absorption (ORCA) protocol \cite{kaczmarek2017a}.

In this work, we focus on a hybrid quantum memory scheme that uses single-photon sources \cite{Reimer2019}, such as quantum dots \cite{Thomas2024}, solid state emitters \cite{Meng2024} or heralded single photons from spontaneous parametric down conversion \cite{Kaneda2016}, together with an atom-like memory system based on negatively charged group-IV color centers in diamond (G4V) \cite{thiering_ab_2018}. The latter is integrated into a single-sided nanophotonic cavity, specifically the Sawfish cavity due to its high coupling efficiency \cite{bopp_sawfish_2024,pregnolato_fabrication_2024}. Tin-vacancy centers (SnVs) are a particularly suitable choice for such quantum memories \cite{Bhaskar2020,Knaut2024,chen_scalable_2024,parker_diamond_2024} when integrated into nanophotonic structures. Due to their reduced sensitivity to charge noise \cite{Bradac2019}, their optical coherence properties are largely preserved. By using state‑dependent reflection from the cavity and implementing high-fidelity fractional quantum gates through a pulse train of optical $\pi/8$ pulses, our specific memory design can achieve gate fidelities exceeding 99\% depending on the bandwidth. We also consider microwave control for the memory scheme, which extends the viability to longer storage times, at the cost of reduced operational rates \cite{pieplow_efficient_2024}.

The goal of this work is to present the technical background for describing the device within a standardized software framework for quantum memory architectures \cite{code_strocka_qm}. The software module we developed allows users to compute Kraus-operators for the read-in and read-out processes using the SiV$^{-}$ or SnV$^{-}$ as a quantum memory. The user can specify optical (limited for now to the SnV) or microwave control, the photon-generation fidelity, bandwidth for read-in and read-out, and the nanophotonic system’s temperature. When microwave control is selected, users may also adjust strain as well as DC and AC magnetic field strengths and orientations as well as the respective defect center, i.e. the SiV$^{-}$ or SnV$^-$. If the user measures data for the spin gate or the cavity, these can be used to estimate the corresponding read-in and read-out Kraus-operators. To support this module, we provide a detailed description of the model for storing and retrieving a time-bin photonic qubit in a spin state, which serves as the basis of the software component.

To model the spin–photon interface, we use closed-form integral expressions covering detailed modeling of spin-photon interaction for the read-in and read-out process, accounting for a broadband photon source and imperfect spin rotations, which are generated by optical control \cite{strocka_token_2025,pieplow_deterministic_2023} or by microwave control \cite{pieplow_efficient_2024}, depending on the color center’s temperature and strain.

The standardized description of the memory requires the use of Kraus-operators \cite{johnston}, which we derive for both: the read-in and read-out process. Such a Kraus representation of the quantum memory system offers the advantage of seamless integration for benchmarking quantum repeater performance and various quantum information applications making it an ideal technical contribution for a large quantum memory simulation software. In \cite{code_strocka_qm} we provide the corresponding module.

Sec. \ref{sec:scheme} introduces our proposed hybrid quantum memory scheme consisting of a single-photon source and a group-IV color center in diamond. Sec. \ref{sec:maths} outlines the mathematical principles concerning the described system. Sec. \ref{sec:numerics} demonstrates several use cases of the developed software. Finally, Sec. \ref{sec:dis} summarizes the results and provides an outlook.
\section{Quantum Memory Scheme}\label{sec:scheme}
The quantum memory scheme is illustrated in Figure \ref{Figure:scheme}.
A single-photon source (SPS 1) emits photons that are transmitted through a fiber to a quantum memory, implemented as a highly efficient single-sided Sawfish crystal cavity containing an SiV$^{-}$ or SnV$^{-}$ center \cite{bopp_sawfish_2024,strocka_token_2025} (see Figure~\ref{Figure:scheme}b). We chose the sawfish cavity as an example due to it large simulated efficiencies \cite{bopp_sawfish_2024} exceeding $98\%$ efficiency of emitter to fiber coupling, but in principle any other half sided cavity spin system is admissible. These photons are entangled with the spin of the color center incorporated into the cavity and subsequently the photon is measured in the $X$-basis, completing the read-in process. For the read-out an extra single-photon source (SPS 2) is required. The respective photon from the SPS interacts with the spin-cavity system by combining the reflection scheme with a measurement in the $Z$-basis of the spin. The reflection scheme for the read-in and read-out process contains an early reflection, a $\pi/2$ rotation, a late reflection and a measurement as illustrated in Figure \ref{Figure:scheme}a,c. 
The reflection steps realize a controlled phase gate which can be optimized by tuning the cavity loss rate $\kappa$, mode frequency $\omega_c$ and central frequency of the emitter $\omega_0$ depending on the bandwidth $\gamma$ of the photon. The spin rotation relies on coherent control of the electron spin of the G4V. Control can be implemented using optical \cite{strocka_token_2025} or microwave control \cite{pieplow_efficient_2024}. Optical control is performed using a Raman scheme \cite{strocka_token_2025} applying a train of optical $\pi/8$ pulses to achieve a high fidelity $\pi/2$ rotation in the presence of dissipative photonic and phononic processes. Notably, both microwave and optical spin control have been experimentally demonstrated for negatively charged silicon- and tin-vacancy (SiV$^{-}$/SnV$^{-}$) center in diamond \cite{Sukachev2017,Becker2016,Rosenthal2023,Debroux2021}, highlighting the versatility of these platforms for quantum information applications. The subsequent $X$-basis measurement and thereby the overall saved state after the read-in process is affected by an important set of the relevant error channels: broadband photons and control imperfections due to photonic and phononic processes, which have an effect on the SiV$^{-}$/SnV$^{-}$ \cite{strocka_token_2025,pieplow_efficient_2024}. Beyond \cite{omlor_entanglement_2024}, we also account for crosstalk during the spin–photon interaction, originating from coupling of the cavity mode to off-resonant optical transitions (see Appendix, Sec. 1 for details).

The read-out process uses an additional single-photon source (SPS 2) whose photon gets entangled with the spin using the reflection scheme in combination with a subsequent $Z$-measurement of the spin.
In this work, we make the following key assumptions, which are fixed parameters in the memory software implementation.

\textit{Single-photon source:} 
We assume that the SPSs photons have a fixed bandwidth and model the spin system as a four-level system where the states are $\ket{1},\ket{2},\ket{A},\ket{B}$. We assume an incoming photonic qubit $\ket{\psi}=\alpha\ket{E}+\beta\ket{L}$ for some $(\alpha,\beta)$ with $\vert\alpha\vert^2+\vert\beta\vert^2=1$ and bandwidth $\gamma_{\rm in}$. For the  additional SPS we assume finite bandwidth photons with $\alpha=\beta=1/\sqrt{2}$, fidelity $F_{\rm aux}$ and bandwidth $\gamma_{\rm out}$. 

\textit{Magnetic field:} For the presented example calculations, we assume static magnetic field strengths $B_{\rm dc}=0.3,1.0,3.0$ T and respective magnetic field orientations $\theta_{\rm dc}=81.4\deg\,,64.62\deg\,,43.11\deg$ from \cite{strocka_token_2025} for optical control of the SnV's spin qubit. In future iterations of our package we will include variable field strength and angle as well as an optical control feature for the SiV. For microwave control the static magnetic field strength $B_{\rm dc}$ as well as the magnitude of the microwave control field $B_{\rm ac}$ and their polar angles $\theta_{\rm dc}$ and $\theta_{\rm ac}$ are variable. We assume the azimuthal angles $\phi_{\rm dc}=0$ and $\phi_{\rm ac}=-\pi/2$ for achieving a rotation around the $y$-axis in the Bloch-sphere \cite{pieplow_efficient_2024} for both the SiV$^{-}$ and SnV$^{-}$ (the orientations are visualized in Figure \ref{Figure:scheme}b.)

\textit{Strain:} The SnV$^{-}$ is in a low-strain environment for optical control \cite{strocka_token_2025}. The SiV$^{-}$/SnV$^{-}$ is placed in a strained environment for microwave control, with the strain contribution given by $E_x = \epsilon_{xx} - \epsilon_{yy}$, where $\epsilon_{xx}$ and $\epsilon_{yy}$ are the normal strain components and $\epsilon_{xy}$ is the shear strain \cite{pieplow_efficient_2024} component. The strain components can be independently adjusted to implement microwave control. 
The strain Hamiltonian is parameterized by a set of constants~\cite{pieplow_efficient_2024}, 
which in the software module are represented by 
$(\alpha_g, \alpha_u, \beta_g, \beta_u)$. 
They are related to the physical strain components according to
\begin{equation}
(\alpha_g, \alpha_u, \beta_g, \beta_u) = 
(d_g E_x,\, d_e E_x,\, 2 d_g \varepsilon_{xy},\, 2 d_e \varepsilon_{xy}) \,,
\end{equation}
where $d_g$ and $d_e$ are the ground- and excited-state strain coupling coefficients.

\textit{Temperature:} The nanophotonic system including the SiV$^{-}$/SnV$^{-}$ is assumed to be cryogenically cooled to temperatures ranging from $100$ mK to $4$ K.

\textit{Power:} For optical transitions, we normalize the laser powers by the saturation power 
and express them as the dimensionless ratio \cite{Podlecki2021}
\begin{align}
    s_i=\frac{2\Omega_i^2}{\gamma^2_{iA}+\Delta^2_i}
\end{align}
where $\Omega_i=\mu_i E_i/\hbar$ is the Rabi-frequency with $E_i$ the amplitude of laser $i$, $\mu_i$ is the transition dipole moment coupling the levels $\ket{i}$ and $\ket{A}$ \cite{strocka_token_2025}, $\gamma_{iA}$ is the spontaneous emission rate using Fermi's golden rule from state $\ket{A}$ to $\ket{i}$ \cite{pieplow_deterministic_2023} and $\Delta_i$ is the detuning between the laser pulse $i$ and the excited state $\ket{A}$ \cite{strocka_token_2025}.
The microwave power is 
\begin{align}
    P^{\rm MW}=B_{\rm ac}^2/2\mu_0 A^{\rm MW},\quad A^{\rm MW}=(\lambda^{\rm MW}/2n)^2
\end{align}
where $B_{\rm ac}$ is the magnetic field strength of the driving field, $A^{\rm MW}$ the effective area with $\lambda^{\rm MW}$ the wavelength of the microwave pulse, $n$ the refractive index of diamond and $\mu_0$ the magnetic susceptibility. 

\textit{Cooperativity:} The cooperativity $C$ is a dimensionless parameter that quantifies the strength of light–matter interaction in a cavity QED system. It is defined as \cite{reiserer_cavity-based_2015}
\begin{align}
C = \frac{g^2}{2\kappa \gamma_a},
\end{align}
where $g$ is the single-photon coupling strength between the emitter and the cavity mode, $\kappa$ is the cavity field decay rate, and $\gamma_a$ is the atomic decay rate calculated from Fermi’s golden rule. Both $2\kappa$ and $\gamma_a$ are given as full widths at half maximum (FWHM) of the respective resonances.
Since achieved cooperativities for SiV$^-$ cavities are limited to $25$ \cite{bhaskar_experimental_2020}, and SnV$^-$ cavities reach only about $1.7$ \cite{Herrmann2024}, the user can define an upper bound for the cooperativity, which we denote by $C_{\rm max}$, defined by the enhancement of the transition from $\ket{A}$ to $\ket{1}$. Notably, the cooperativity between $\ket{1}$ and $\ket{A}$ is approximately as large as the cooperativity between $\ket{2}$ and $\ket{B}$ for G4Vs \cite{omlor_entanglement_2024}.

Our software package optimizes under cooperativity constraints. In addition, it allows users to specify their reported cavity parameters, since fabricated cavities often deviate from the optimized design. It can thus provide local optima useful for fabrication targets and estimate quantum memory performance for arbitrary cavity parameters and photon central frequencies.

\textit{Spin gate}: For both microwave and optical control, a spin $\pi/2$ rotation is simulated if the user chooses to generate an optimized spin rotation. However, users may also specify the spin gate manually, for example when they are incorporating the routine into their own numerical simulations or interfacing it with experimental control hardware.
 In our module, users are required to provide the propagated states assuming the initial states
\begin{align}
\rho_{1}=\ket{1}\bra{1},\quad\rho_{2}=\ket{2}\bra{2},\quad\rho_+=\ket{+}\bra{+},\quad \rho_{R}=\ket{R}\bra{R}
\end{align}
with $\ket{+}=\frac{1}{\sqrt{2}}(\ket{1}+\ket{2})$ and $\ket{R}=\frac{1}{\sqrt{2}}(\ket{1}+{\rm i}\ket{2})$.

\textit{Time:} The total processing time $T_1$ is given by
\begin{align}
    T_1&=T_{\rm read-in}+T_{\rm read-out}+T_s,\\
    T_k&=T_{\rm tb, k} + T_g + T_m + T_{c,\rm k}
\end{align}
for $k=\rm read-in,read-out$ where $T_{\rm tb,k}$ is the time allocated for a time-bin qubit, $T_g$ is the control gate duration, $T_m$ is the measurement time, $T_s$ is the storage time, and $T_{c,k}$ is the transmission time for the respective process. We set $T_{\rm tb,k} = 20 T_{\rm lt,k}$, where $T_{\rm lt,k}$ is the lifetime of the single-photon source used for the read-in or read-out, respectively, and choose $T_g = 40\sigma$ for optical control with 
$
\sigma = \tau_{\pi/8}/2\sqrt{2\ln(2)}
$,
where $\tau_{\pi/8}$ is the full-width at half maximum of the optimized $\pi/8$ pulse for the respective magnetic field strength listed in \cite{strocka_token_2025}. We take $T_m = 100$ ps \cite{cherednichenko_low_2021}, corresponding to the dead time of the photon detectors \cite{Grotowski2025}. The communication time is given by $T_{c,k} = L_k / c$, where $L_k$ is the fiber length for read-in or read-out, respectively and $c$ is the speed of light in the fiber.

\section{Modeling}\label{sec:maths}
We outline the steps to compute Kraus-operators that map the photonic state to the spin state during the read-in process. We assume the spin is initialized in the state $\ket{1}$. We then compute Kraus-operators mapping the stored spin state back to the photonic qubit for the read-out process.

\subsection{Single-Photon Source}

Photons from single-photon sources can be modeled in time-bin encoding with the state space
\begin{align}
    \mathcal{H}={\rm span}\{\ket{E},\ket{L}\}.
\end{align}
We define the time-bin qubits as superpositions of single photons $\ket{\omega_{E,L}}$ in frequency space so that
\begin{align}
    \ket{E,L}=\int_\mathbb{R} S(\omega-\omega_0)\ket{\omega_{E,L}}\,{\rm d}\omega
\end{align}
where $S(\omega)$ is the emission spectrum with a central frequency $\omega_0$. We assume a Lorentzian intensity profile \cite{tran_nanodiamonds_2017}, i.e.
\begin{align}
    S(\omega)=\mathcal{N}\frac{\epsilon_0}{{\rm i}\omega+\gamma/2}
\end{align}
with the normalization factor $\mathcal{N}=\left(\int_\mathbb{R} \vert S(\omega)\vert^2\right)^{-1/2}$, a field strength $\epsilon_0$ and the bandwidth $\gamma=\gamma_{\rm in},\gamma_{\rm out}$ where $\gamma_{\rm in,out}$ denote the bandwidth of the incoming photon for the read-in and read-out process, respectively.
An arbitrary photonic qubit is then given by
\begin{align}
    \ket{\psi_{\rm ph}}=\alpha\ket{E}+\beta\ket{L}
\end{align}
with $\alpha,\beta\in\mathbb{C}$ and $\vert\alpha\vert^2+\vert\beta\vert^2=1$.
We assume that the emitted photon gets depolarized during its generation according to the quantum channel \cite{tiurev2021fidelity}
\begin{align}\label{eq:photon}
    \mathcal{E}(\rho_{\rm ph})=(1-\epsilon)\rho_{\rm ph}+\epsilon{\rm tr}(\rho_{\rm ph})\frac{\mathds{1}}{2}
\end{align}
with $\epsilon=2(1-F_{\rm ph})$ and $\rho_{\rm ph}=\ket{\psi_{\rm ph}}\bra{\psi_{\rm ph}}$
where the quantity $F_{\rm ph}$ denotes the state fidelity between the states $\rho_{\rm ph}$ and $\mathcal{E}(\rho_{\rm ph})$ \cite{Liang2019}. For the read-in process we do not include depolarization during photon generation because it is more intuitive for the user to depolarize a photonic qubit according to Eq. \eqref{eq:photon} and subsequently apply the Kraus map modeling the read-in process. For read-out the photon generation fidelity is called $F_{\rm aux}$ and it holds $\alpha=\beta=1/\sqrt{2}$.

\subsection{Cavity Design}
To store a broadband photon in the cavity coupled spin we must carefully engineer the cavity such that the controlled phase gate fidelity between the spin and photon \cite{Liang2019}
\begin{align}\label{eq:fidd}
    F_{\rm sp}=\bra{\psi}\rho_{\rm sp}\ket{\psi}
\end{align}
with $\ket{\psi}=\frac{1}{\sqrt{2}}(\ket{1}+\ket{2})$ is maximized (see Appendix, Sec. 2 for details). To evaluate the fidelity given in Eq. \eqref{eq:fidd}, the spin-dependent reflection coefficients are required. We assume that a single cavity mode is coupled to the $\ket{1} \leftrightarrow \ket{A}$ and $\ket{2} \leftrightarrow \ket{B}$ transitions. The reflection coefficients are given by \cite{reiserer_cavity-based_2015,omlor_entanglement_2024}
\begin{align}\label{eq:refl_coeff}
    R_{1}(\omega)&=-1+\frac{2 \kappa({\rm i}\Delta_{1A}+\gamma_{1A}/2)}{({\rm i}\Delta_c+\kappa)({\rm i}\Delta_{1A}+\gamma_{1A}/2)+\vert g_{1A}\vert^2},\\
    R_{2}(\omega)&=-1+\frac{2 \kappa ({\rm i}\Delta_{2B}+\gamma_{2B}/2)}{({\rm i}\Delta_c+\kappa)({\rm i}\Delta_{2B}+\gamma_{2B}/2)+\vert g_{2B}\vert^2}
\end{align}
with the detunings $\Delta_{1A}:=\omega-\omega_{1A}, \Delta_{2B}=\omega-\omega_{2B}$, the transition frequencies $\omega_{1A},\omega_{2B}$, the atomic decay rates $\gamma_{1A},\gamma_{2B}$, the detuning $\Delta_c=\omega-\omega_c$ with the cavity mode central frequency of the SiV$^{-}$/SnV$^{-}$ $\omega_c$ and the coupling strength between the SiV$^{-}$/SnV$^{-}$ and the photonic mode in the cavity
\begin{align}
    g_{1A}={\rm i}\sqrt{\frac{\omega_c}{2\hbar\epsilon_0\epsilon_r 
    V}}\bra{1}\bm{\epsilon}\cdot\mathbf{d}\ket{A},\label{eq:coupling1}\\
    g_{2B}={\rm i}\sqrt{\frac{\omega_c}{2\hbar\epsilon_0\epsilon_r\label{eq:coupling2} 
    V}}\bra{2}\bm{\epsilon}\cdot\mathbf{d}\ket{B}~,
\end{align}
where the mode volume is $V=V_{\rm eff}\frac{\lambda^3}{2n^3}$ \cite{bhaskar_experimental_2020}, the wavelength of the cavity mode
$\lambda=2\pi c/\omega_{c}$, the speed of light $c$ and the mode orientation $\bm{\epsilon}$. We assume the effective mode volume $V_{\rm eff}=1.8$ and mode orientation $\bm{\epsilon}=\bm{e_z}$, i.e. the mode is parallel to the SiV's/SnV's symmetry axis.

To maximize the fidelity shown in Eq. (\ref{eq:fidd}) the software package can optimize the triple $(\omega_0,\omega_c,\kappa)$ using simplicial homology global optimization \cite{endres_simplicial_2018, 2020SciPy-NMeth}.
Optimized values for $(\omega_c,\kappa)$ yield cooperativities $\mathbf{C}=(C_{1A},C_{2A},C_{1B},C_{2B})$, calculated as $C_{kl}=2|g_{kl}|^2/(\kappa\gamma_{kl})$ for $k=1,2$ and $l=A,B$. Care must be taken with the factor of two, since $\kappa$ denotes the half-width at half-maximum (HWHM) of the cavity mode, while $\gamma_{kl}$ refers to the full width at half-maximum (FWHM) of the optical transition. Accordingly, the cooperativity definition consistent with HWHM values should be used when calculating the Purcell enhancement \cite{Wang2021}.

\subsection{Spin Rotation}

The spin $\pi/2$ rotation is an essential step for the quantum memory scheme. It can be performed using optical or microwave control. Depending on the choice of the control the state space is $\mathcal{L^{\rm opt}}$ for optical control and $\mathcal{L^{\rm mw}}$ for microwave control. They are defined by
\begin{align}
    &\mathcal{L^{\rm opt}}=\{\ket{i}\bra{j}\vert i,j=1,...,8\},\\
    &\mathcal{L^{\rm mw}}=\{\ket{i}\bra{j}\vert i,j=1,...,4\}.
\end{align}

\textit{Raman control:} Our software package propagates $\rho_{ij,0}=\ket{i}\bra{j}$ for $i,j=1,2$ subject to the Lindblad-master equation, i.e. \cite{pieplow_deterministic_2023,strocka_token_2025} 
\begin{align}\label{eq:lind}
    \dot{\rho}&=-{\rm i}[H(t),\rho]+\mathcal{L}_{D}^{(T,\mathbf{C},\bm{B_{\rm dc}})}(\rho),\\
    \rho(0)&=\rho_{0,ij}
\end{align}
with the Hamiltonian
\begin{align}
   H(t)=H_0-\bm{\mu}\cdot\mathbf{E}(t),
\end{align}
and the dissipator
\begin{align}
    \mathcal{L}_{D}^{(T,\mathbf{C},\bm{B_{\rm dc}})}(\rho)=\sum_{k}\gamma_k \left( L_{k}\rho(t) L_{k}^\dagger-\frac{1}{2}\{L_{k}^\dagger L_{k},\rho(t)\}\right)
\end{align}
with the free contribution $H_0$ and the light-matter interaction term $\bm{\mu}\cdot\mathbf{E}(t)$ using the optimized parameters for $B_{\rm dc}=0.3,1.0,3.0$ T and $\theta_{\rm dc}$ taken from \cite{strocka_token_2025} for the SnV. The detailed description of the Hamiltonian $H^{\rm opt}(t)$ and temperature, cooperativity and magnetic field dependent relaxation rates $\gamma_k$ as well as Lindblad-operators $\{L_k\}$ modeling photonic decay processes using Fermi's golden rule and fast phononic processes are explained in \cite{pieplow_deterministic_2023,strocka_token_2025}.

\textit{Microwave control:} We simulate the $\pi/2$ rotation using microwave control using the Lindblad-master equation shown in Eq. (\ref{eq:lind}) with the Hamiltonian \cite{pieplow_efficient_2024}
\begin{align}
    H(t)=H_{\rm dc}+H_{\rm ac}(t)
\end{align}
consisting of the time-invariant part $H_{\rm dc}$ due to the DC magnetic field and the driving term $H_{\rm ac}(t)$ modeling an oscillating AC magnetic field and the dissipator $\mathcal{L}_D^{(T,E_x,\epsilon_{xy},\bm{B_{\rm dc}})}(\rho)$ containing only the temperature, strain and field $\bm{B_{\rm dc}}$ dependent phononic processes. The $\pi/2$ rotation is achieved by driving the AC field for a quarter of a full Rabi oscillation for both the SiV$^{-}$ and SnV$^{-}$ \cite{pieplow_efficient_2024}.  

\subsection{Kraus-Operators}

To model the quantum channel for the read-in and read-out we explain the formalism for computing Kraus-operators below. 
To model a quantum channel $\mathcal{D}$ with Kraus-operators $\{K_m\}$ with $\sum_{m=1}^4 K_m^\dagger K_m\le \mathds{1}$ \cite{Bongioanni2010,Bhandari2016} such that 
\begin{align}\label{eq:evol}
    \rho^{ij}(T)=\sum_{m=1}^4 K_m \ket{i}\bra{j} K_m^\dagger,\quad i,j=1,2
\end{align}
we apply the Choi-Jamiołkowski-Isomorphism \cite{johnston}. The formalism requires the Choi matrix which is of the form
\begin{align}
    J=\sum_{i,j=1}^2\mathcal{D}(\ket{i}\bra{j})\otimes \ket{i}\bra{j}.
\end{align}
The Kraus-operators are
\begin{align}\label{eq:kraus}
    K_m=\sqrt{\lambda_m}\begin{pmatrix}
        \langle 1\vert\psi_m\rangle & \langle 2\vert\psi_m\rangle\\
        \langle 3\vert\psi_m\rangle & \langle 4\vert\psi_m\rangle
    \end{pmatrix}
\end{align}
with eigenvalues $\{\lambda_m\}$ and eigenvectors $\{\ket{\psi_m}\}$ of the Choi matrix $J$. The Kraus map is not trace-preserving because the read-in and read-out process require a measurement. 

To evaluate Kraus-operators for the read-in and read-out process we need to propagate the full basis of photonic - and spin states subject to the quantum channel for the read-in and read-out process, respectively. A detailed mathematical exploration of the spin states after measurement including imperfect spin gates using microwave or optical control encountering photonic and fast phononic processes and a detailed spin-photon interaction model including crosstalk which goes beyond the work in \cite{omlor_entanglement_2024} is outlined in the Appendix, Sec. 1. Let's assume a photonic qubit $\rho_{\rm ph}$ is passing all reflection steps including the $X$-measurement resulting in the measurement outcomes $\rho_+$ and $\rho_-$.  

To compute Kraus-operators for the read-in process we evaluate
\begin{align}
    \mathcal{D}_{\rm read-in,+}(\ket{i}\bra{j})=\rho_{\rm sp,+}^{ij}(T_{\rm read-in}),\quad i,j=E,L,\\
    \mathcal{D}_{\rm read-in,-}(\ket{i}\bra{j})=\rho_{\rm sp,-}^{ij}(T_{\rm read-in}),\quad i,j=E,L
\end{align}
with the quantum channel for the read-in $\mathcal{D}_{\rm read-in,+/-}$ assuming a measurement in $\ket{+/-}$ for the photonic qubit.
To compute Kraus-operators for the read-out process we evaluate
\begin{align}
   \mathcal{D}_{\rm read-out,1}(\ket{i}\bra{j})=\rho_{\rm ph,1}^{ij}(T_{\rm read-out}),\quad i,j=1,2,\\
   \mathcal{D}_{\rm read-out,2}(\ket{i}\bra{j})=\rho_{\rm ph,2}^{ij}(T_{\rm read-out}),\quad i,j=1,2
\end{align}
with the quantum channel for the read-out $\mathcal{D}_{\rm read-out,1/2}$ assuming a measurement in the state $\ket{1/2}$ for the spin qubit. Details regarding the read-out process are given in the Appendix, Sec. 4. We compute Kraus-operators by evaluating Eq. (\ref{eq:kraus}). For a photonic qubit of the form shown in Eq. (\ref{eq:photon}) the spin state after read-in is 
\begin{align}
    \rho_{\rm sp,+/-}=\frac{\mathcal{D}_{\rm read-in,+/-}(\rho_{\rm ph})}{\eta_{+/-}}
\end{align}
with $\eta_{+/-}={\rm tr}(\mathcal{D}_{\rm read-in,+/-}(\rho_{\rm ph}))$ describing the trace reduction of the read-in process depending on the measurement $\ket{+}$ or $\ket{-}$.
After read-out we get
\begin{align}
    \rho_{\rm ph,+/-,1/2}=\frac{\mathcal{D}_{\rm read-out,1/2}(\rho_{\rm sp,+/-})}{\eta_{+/-,1/2}}.
\end{align}
with $\eta_{+/-,1/2}={\rm tr}(\mathcal{D}_{\rm read-out,1/2}(\rho_{\rm sp,+/-}))$ representing the trace reduction for the read-out procedure depending on whether the input state was measured in $\ket{+}$ or $\ket{-}$ respectively. To evaluate the quantum memory performance using Kraus-operators, we define an average fidelity and trace reduction that account for both the read-in and read-out processes across the Bloch sphere of photonic qubits (see Appendix, Sec. 5 for details).
\subsection{Approximation Error}
Consider the initial state $\rho_0=\ket{1}\bra{1}$ and the propagated state subject to the $\pi/8$ rotation at final time $T$ which is called $\rho(T)$. Now define $\tilde{\rho}(T)=\sum_{i,j} \tilde{\rho}_{ij}(T)\ket{i}\bra{j}$ with
\begin{align}
    \tilde{\rho}_{ij}(T)=\begin{cases}
        \rho_{ij}(T),\quad{\rm if}\quad i,j=1,2,\\
        0,\quad{\rm else}
    \end{cases}.
\end{align}
Subsequently, the approximation error is
\begin{align}\label{eq:error}
    e=\norm{\rho(T)-\tilde{\rho(T)}}_1
\end{align}
where $\norm{A}_1=\max_{1 \leq j \leq 8} \left( \sum_{i=1}^8 |A_{ij}| \right)$ denotes the 1-norm of a matrix.
We also use Eq. \eqref{eq:error} for evaluating the approximation error for microwave control. As we will show in Section \ref{sec:numerics}, for errors smaller than $10^{-4}$, the resulting trace reduction can be considered negligible.
\section{Numerical Examples}\label{sec:numerics}
\begin{figure*}[tb]
    \centering
    \includegraphics[width=\linewidth]{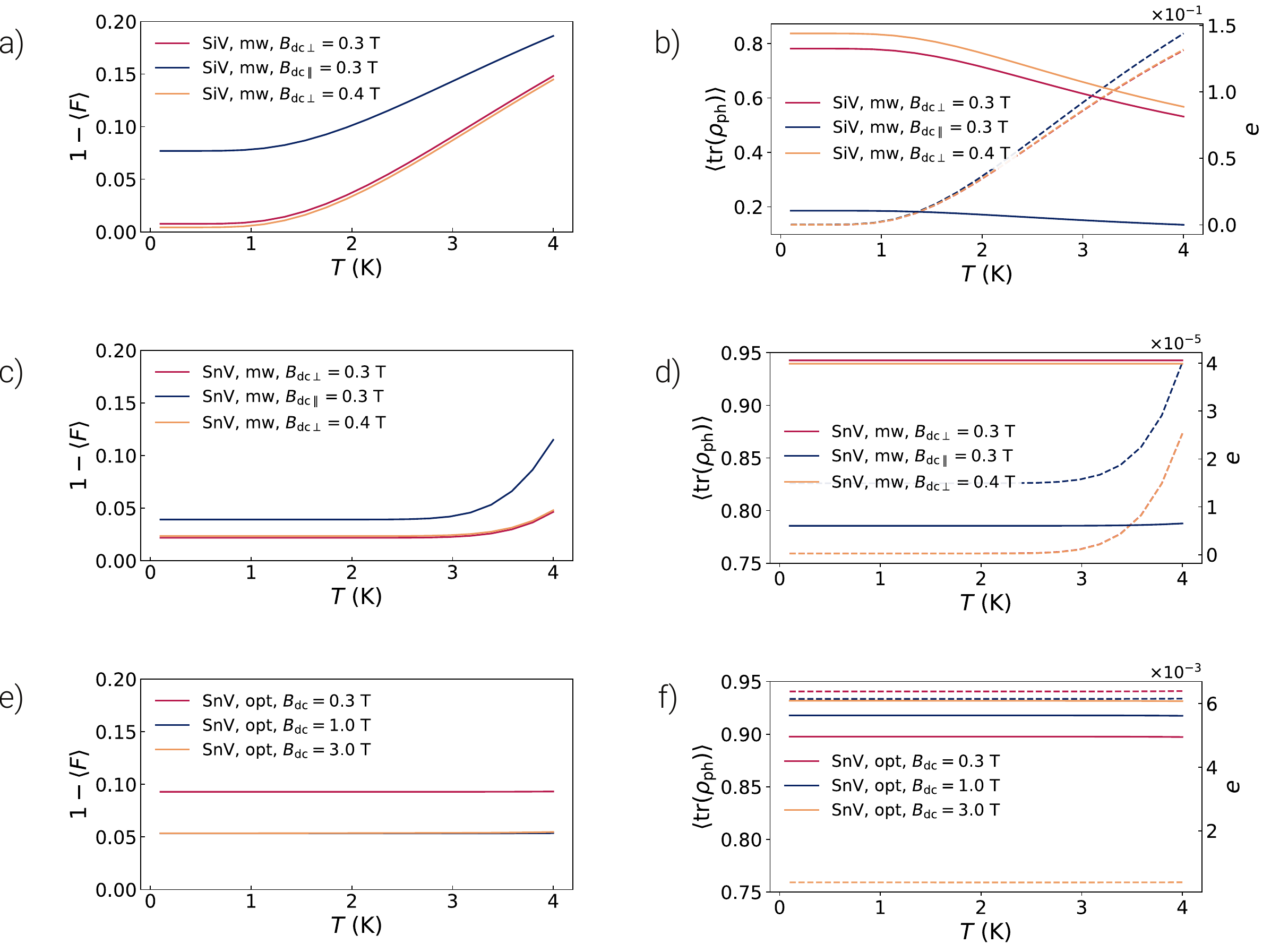}
    \caption{Visualization of the infidelity $1-\langle F\rangle$ and trace $\langle {\rm tr}(\rho_{\rm ph})\rangle$ of the photonic qubit after read-out (see Appendix, Sec. 5 for details) as a function of the system's nanophotonic temperature $T$ assuming the bandwidth of the photonic qubit for the read-in and read-out process is $\gamma_{\rm in,out}=0.1$ GHz, the AC magnetic field for microwave control of the SiV$^{-}$ (SnV$^{-}$) is $B_{\rm ac}=3.7 (1.0)$ mT, the axial strain is $E_x=4\cdot 10^{-5} (0)$ and the shear strain is $\epsilon_{xy}=0 (0)$ for the SiV$^{-}$ (SnV$^{-}$), the fidelity of the auxiliary photon source for read-out is $F_{\rm aux}=1.0$ for all the considered cases and we limit the cooperativity to $C_{\rm max}=12.5$.}
    \label{Figure:temp}
\end{figure*}
\begin{figure*}[tb]
    \centering
    \includegraphics[width=\linewidth]{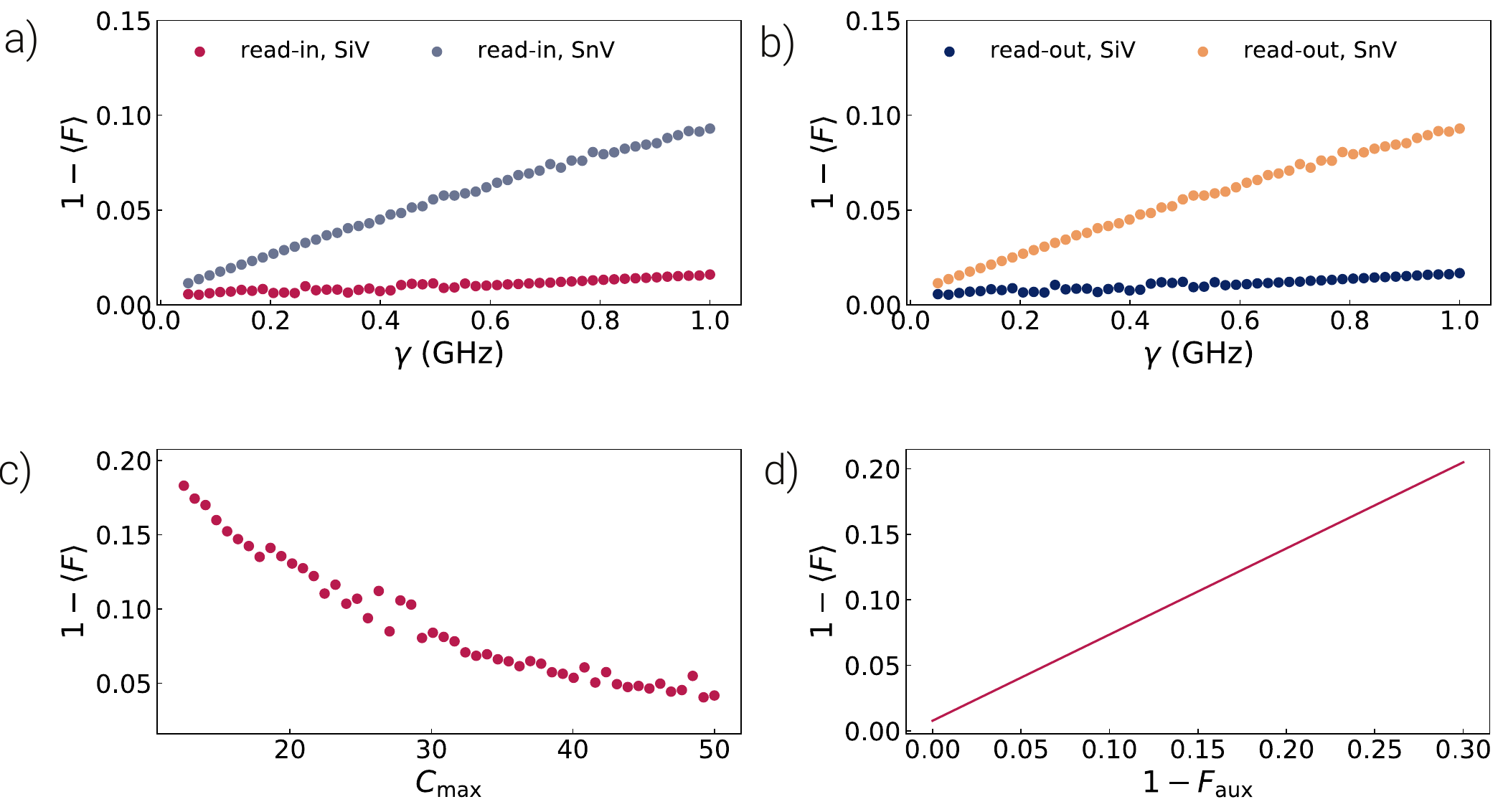}
    \caption{a) Visualization of the infidelity as a function of the bandwidth for the SiV$^{-}$ and SnV$^{-}$ for the read-in and b) read-out process. For the SnV$^{-}$ the trace is ${\rm tr}(\rho_{\rm ph})\approx 0.94$ and for the SiV$^{-}$ it is $\langle {\rm tr}(\rho_{\rm ph})\rangle= 0.80$ on the considered range. c) Visualization of the infidelity as a function of the maximal allowed cooperativity assuming all microwave control for the SnV$^{-}$ with $B_{\rm dc}=0.3$ T, $B_{\rm ac}=1$ mT in a low strain environment at $T=0.1$ K for $\gamma_{\rm in,out}=1$ GHz. The average trace reduction on that range is $\langle {\rm tr}(\rho_{\rm ph})\rangle\approx 0.9580$. Due to a rather noisy dependence of the trace reduction on the cooperativity we omit that graph here. For the SiV$^{-}$ the optimal cooperativity is below $12.5$ leading to no improvement on the considered range which is why we omit the SiV$^{-}$ representation here. d) Visualization of the infidelity as a function of the infidelity of the auxiliary photon source for read-out considering the SiV$^{-}$ as defect center. The trace is $\langle {\rm tr}(\rho_{\rm ph})\rangle=0.7812$ on the considered range.}
    \label{Figure:sweeps}
\end{figure*}
The main objective of the software module \cite{code_strocka_qm} is the calculation of Kraus operators that describe the action of the memory on a photonic qubit after storage and retrieval. We now provide examples of how to use the module based on the computation of a Kraus representation of the memorization procedure. As an instructive example, we calculate the average fidelity for reading in and reading out a photonic qubit (details are shown in the Appendix, Sec. 5). We also show the trace of the photonic qubit after read-out to quantify loss and decoherence in the system.

We visualize the infidelity and trace as a function of the system's nanophotonic temperature from $100$ mK to $4$ K, which is the experimentally relevant temperature range for control experiments \cite{Sukachev2017,Becker2016,Rosenthal2023,Debroux2021}. To cover the broad range of accessible magnetic field strengths \cite{Sukachev2017, Rosenthal2023}, we perform simulations for different fields and for both the SiV and SnV. In addition, we sweep the bandwidth of the single-photon sources to assess how source spectral properties impact overall performance. Since cavity cooperativity is fabrication-limited \cite{Herrmann2024}, we also evaluate the influence of the cooperativity bound. Finally, we analyze the memory performance as a function of the fidelity of the auxiliary photon source used for read-out, which is essential for determining the minimum source quality required for practical implementations. As a last example, we show how user-defined cavity parameters can be used to predict the performance of a G4V based quantum memory integrated into a cavity.  

In Figure \ref{Figure:temp} we visualize the infidelity and trace as a function of the temperature for the SiV$^{-}$ and SnV$^{-}$ using three magnetic field strengths and show the results for both microwave and optical control to produce the spin $\pi/2$ rotation. For the simulations we assume a range of experimentally relevant parameters \cite{Dietrich2020,Sukachev2017, Rosenthal2023}: $\gamma_{\rm in,out}=0.1$ GHz, $B_{\rm ac}=1.0$ mT $(3.7)$ mT for the SiV$^{-}$ (SnV$^-$), strain parameters $E_x=4\cdot 10^{-5}$ $(0)$, $\epsilon_{xy}=0$ $(0)$ and the upper bound for the cooperativity $C_{\rm max}=12.5$. The chosen value $C_{\mathrm{max}} = 12.5$ does not limit the experimentally achievable fidelity for the SiV$^-$ \cite{strocka_repeater_2025}, but it does constrain the fidelity for the SnV$^-$, consistent with the cooperativities reported for SnV$^{-}$ cavities in~\cite{Herrmann2024}. It becomes apparent that the infidelity increases and the trace decreases as the temperature rises. The infidelity rise is due to phonon induced spin dephasing \cite{pieplow_efficient_2024}. Since phononic excitation also plays an increasingly important role as the temperature rises the population leaves the logical subspace leading to a rise of the approximation error and trace drop. To justify the estimated approximation error bound of $10^{-4}$ for negligible trace reduction, we refer to Fig.~\ref{Figure:temp}d). 
In this figure, the approximation error remains below $4\times10^{-5}$, while the trace shows no noticeable variation on the plotted scale. 

In Fig.~\ref{Figure:temp}a) and c) we observe that the infidelity is smaller when the DC magnetic field is oriented orthogonally to the symmetry axis than when it is applied parallel to it. This is expected because the orthogonal orientation yields a smaller optical splitting $\Delta\omega_s := \omega_{2B} - \omega_{1A}$ \cite{strocka_repeater_2025}. Comparing the SiV and SnV in these figures, we further see that the SiV exhibits a lower infidelity than the SnV for temperatures below approximately $2$ K, whereas the trend reverses as the temperature approaches $4$ K. The superior low-temperature performance of the SiV arises from the chosen cooperativity limit, which does not reduce the SiV fidelity but does limit the SnV fidelity \cite{strocka_repeater_2025}. At higher temperatures, however, phonon-induced spin dephasing becomes increasingly significant for the spin gate. The microwave spin gate used for the SiV is more sensitive to this effect than the corresponding gate for the SnV. In Fig.~\ref{Figure:temp}e) we observe that the infidelity remains essentially independent of temperature when optical control is used for the spin rotation, which follows from the short gate duration relative to the timescale of phonon-induced spin dephasing \cite{strocka_token_2025}.

In Figure \ref{Figure:sweeps} we visualize the infidelity as a function of the bandwidth for read-in and read-out as well as the upper bound on the cooperativity for the SnV and the infidelity of the auxiliary photon source.
In Fig.~\ref{Figure:sweeps}a) and b) we plot the infidelity $1-\langle F\rangle$ as a function of the bandwidth $\gamma$ of the incoming and outgoing photonic modes, using either the SiV$^{-}$ or SnV$^{-}$ as the quantum memory. The simulations assume a magnetic field strength $B_{\rm dc}=300$ mT with orientation $\theta_{\rm dc}=\pi/2$, microwave control with $B_{\rm ac}=1$ mT (3.7 mT) and $\theta_{\rm ac}=0$, a temperature of $T=0.1$ K, axial strain $E_x=4\cdot 10^{-5}$ (0), shear strain $\epsilon_{xy}=0$ (0), an auxiliary photon-source fidelity of $F_{\rm aux}=1.0$, and a cooperativity bound of $C_{\rm max}=12.5$. As expected, the infidelity increases with bandwidth. This trend reflects the reduced phase contrast between the transitions $\ket{1}\leftrightarrow\ket{A}$ and $\ket{2}\leftrightarrow\ket{B}$ as the photonic spectral width broadens \cite{omlor_entanglement_2024}.

When comparing Figure \ref{Figure:sweeps}a) and b) we observe that the infidelities for read-in and read-out are equal for the SiV$^-$.
The same behavior becomes apparent for the SnV$^-$. That is comprehensible because both the read-in and read-out process entail the same scheme for producing entanglement, hence the same errors are produced under an equal parameter setting. We also observe that the SiV's infidelity is smaller than the SnV's infidelity on the considered range. That is due to the cooperativity limit $C_{\rm max}=12.5$. While the optimal cooperativity for the SiV$^{-}$ for $\gamma_{\rm in,out}=1$ GHz is below $C_{\rm max}$ it is not for the SnV$^-$. No source of error for the SiV$^{-}$ is produced due to that bound but for the SnV$^{-}$ as it becomes apparent in Figure \ref{Figure:sweeps}c). In the Appendix, Sec. 6 we list microwave and laser powers and total processing times for various parameter settings.

In Figure \ref{Figure:sweeps}c) we illustrate the infidelity $1-\langle F\rangle$ as a function of the upper bound of the cooperativity $C_{\rm max}$ for the optimization of a cavity for the SnV$^{-}$ in a low strain environment at $T=0.1$ K for $\gamma_{\rm in,out}=1$ GHz at the magnetic field strength $B_{\rm dc}=300$ mT and orientation $\theta_{\rm dc}=\pi/2$ using microwave control with $B_{\rm ac}=1$ mT and the orientation $\theta_{\rm ac}=0$. We observe that the infidelity decreases as the cooperativity bound increases. That is the expected behavior since a higher cooperativity yields a flatter shape of the phase contrast in a local environment of the optimized central frequency of the incoming photonic mode \cite{omlor_entanglement_2024}.

In Figure \ref{Figure:sweeps}d) we illustrate the infidelity as a function of the infidelity of the auxiliary photon source for read-out assuming the SiV$^{-}$ as the defect center with $B_{\rm dc}=300$ mT, $B_{\rm ac}=3.7$ mT, $\theta_{\rm dc}=\pi/2$, $\theta_{\rm ac}=0$, $E_x=4\cdot 10^{-5}$, $\epsilon_{xy}=0$ and $\gamma_{\rm in,out}=0.1$ GHz. We observe a linear rise of the infidelity as the infidelity of the photon source increases. That is intuitive because the quality of the auxiliary photon is essential to produce high fidelity read-out. However, we observe $1-\langle F\rangle\approx 0.2$ at $1-F_{\rm aux}=0.3$ which occurs because the error in the photon gets washed out when performing measurement of the spin. 

Finally, we provide an example in which the user supplies experimental cavity data together with the emission frequency of the photonic mode after frequency conversion for the read-in and read-out processes. We consider an SnV$^{-}$ with a DC magnetic field of $B_{\rm dc}=300$ mT oriented orthogonally to its symmetry axis in a low-strain environment, and we assume a photon bandwidth of $\gamma_{\rm in,out}=0.1$ GHz for both operations. Using the software module \cite{code_strocka_qm}, the user first determines an optimized cavity configuration and central photon frequency specified by $(k, \delta_c, \delta_0)$, which then serves as the experimental design target. The software returns the theoretical optimal parameters $\kappa_{\rm theo}=42.85$ GHz, $\delta_{c,\rm theo}=-7.67$ GHz, and $\delta_{0,\rm theo}=-3.41$ GHz. Experimentally, however, the user measures $\kappa_{\rm exp}=41$ GHz, $\delta_{c,\rm exp}=-4$ GHz, and $\delta_{0,\rm exp}=-3$ GHz.
We evaluate the average fidelity and trace reduction on the Bloch sphere (see Appendix, Sec. 5) for both the theoretical and experimental parameter sets. The theoretical values are $\langle F\rangle = 0.9898$ and $\langle {\rm tr}(\rho_{\rm ph})\rangle = 0.9688$, whereas the experimentally inferred values are $\langle F\rangle = 0.8306$ and $\langle {\rm tr}(\rho_{\rm ph})\rangle = 0.9779$. Thus, although the trace remains nearly unchanged, the fidelity decreases by about $0.15$, indicating that the experimental configuration leaves considerable room for improvement to achieve a high-quality spin memory.

The results indicate that the developed software performs reliably across a wide range of parameters, validating its practical utility.

\section{Discussion and Outlook}\label{sec:dis}

Our executable quantum memory component leverages negatively charged silicon- and tin-vacancy (G4V) centers to store photonic time-bin qubits and demonstrates both robustness and versatility. The module computes the Kraus operators describing the read-in and read-out processes by internally optimizing all relevant control pulses and cavity parameters, including the locally optimal optical controls from \cite{strocka_token_2025}. The full implementation is available in \cite{code_strocka_qm}.

Users may select optical control for the SnV$^{-}$ or microwave control for the SiV$^{-}$ and SnV$^{-}$ for both read-in and read-out, and in all cases they can configure photon-generation fidelity of the auxiliary photon source, bandwidth of the incoming photon for read-in and read-out and the nanophotonic system’s temperature. The user can also choose between the optimized phase gate and manually chosen parameters to produce a phase gate as well as the optimized spin gate and manually set propagated spin states. If microwave control is chosen, strain as well as DC and AC magnetic field strengths and orientations are also adjustable; under optical control these parameters remain fixed.

The module can be further extended to enhance its versatility. For example, lifting the current constraints on optical control to permit arbitrary magnetic field strengths would broaden the accessible parameter space. 
Other group-IV vacancy centers, such as the germanium, and lead vacancies \cite{Thiering2018}, can also be included, serving an even broader research community. In future iterations of the software module we plan to include the carbon nuclear spin as the quantum memory to increase the coherence time.

Overall, the software component \cite{code_strocka_qm}, which is openly accessible, will enable a broad user base to study group-IV vacancy centers in a quantum-network setting making it a highly valuable research tool.

\section*{Author Contributions}
Y.S. and G.P. conceptualized the described software module with Y.S. implementing and testing core functionalities. M.B. provided the code specific to microwave spin control. T.S. and G.P developed the core idea and provided overall project supervision. All authors contributed to writing and refining the manuscript.\\

\section*{Conflict of Interest}
The authors declare no financial or commercial conflict of interest.
\section*{Acknowledgements}
We would like to thank Elizabeth Robertson, Fenglei Gu and Johannes Borregaard for fruitful discussions during the development of the model.
Funding for this project was provided by the German Federal Ministry of Education and Research (BMBF, project QPIS, No. 16KISQ032K; project DINOQUANT 13N14921, ERC StG project QUREP of the EC, No. 851810). 

\bibliographystyle{aipnum4-1}
\bibliography{memory_revised.bib}\cleardoublepage

\begin{thebibliography}{79}%
\makeatletter
\providecommand \@ifxundefined [1]{%
 \@ifx{#1\undefined}
}%
\providecommand \@ifnum [1]{%
 \ifnum #1\expandafter \@firstoftwo
 \else \expandafter \@secondoftwo
 \fi
}%
\providecommand \@ifx [1]{%
 \ifx #1\expandafter \@firstoftwo
 \else \expandafter \@secondoftwo
 \fi
}%
\providecommand \natexlab [1]{#1}%
\providecommand \enquote  [1]{``#1''}%
\providecommand \bibnamefont  [1]{#1}%
\providecommand \bibfnamefont [1]{#1}%
\providecommand \citenamefont [1]{#1}%
\providecommand \href@noop [0]{\@secondoftwo}%
\providecommand \href [0]{\begingroup \@sanitize@url \@href}%
\providecommand \@href[1]{\@@startlink{#1}\@@href}%
\providecommand \@@href[1]{\endgroup#1\@@endlink}%
\providecommand \@sanitize@url [0]{\catcode `\\12\catcode `\$12\catcode
  `\&12\catcode `\#12\catcode `\^12\catcode `\_12\catcode `\%12\relax}%
\providecommand \@@startlink[1]{}%
\providecommand \@@endlink[0]{}%
\providecommand \url  [0]{\begingroup\@sanitize@url \@url }%
\providecommand \@url [1]{\endgroup\@href {#1}{\urlprefix }}%
\providecommand \urlprefix  [0]{URL }%
\providecommand \Eprint [0]{\href }%
\providecommand \doibase [0]{http://dx.doi.org/}%
\providecommand \selectlanguage [0]{\@gobble}%
\providecommand \bibinfo  [0]{\@secondoftwo}%
\providecommand \bibfield  [0]{\@secondoftwo}%
\providecommand \translation [1]{[#1]}%
\providecommand \BibitemOpen [0]{}%
\providecommand \bibitemStop [0]{}%
\providecommand \bibitemNoStop [0]{.\EOS\space}%
\providecommand \EOS [0]{\spacefactor3000\relax}%
\providecommand \BibitemShut  [1]{\csname bibitem#1\endcsname}%
\let\auto@bib@innerbib\@empty
\bibitem [{\citenamefont {Bopp}\ \emph {et~al.}(2024)\citenamefont {Bopp},
  \citenamefont {Plock}, \citenamefont {Turan}, \citenamefont {Pieplow},
  \citenamefont {Burger},\ and\ \citenamefont {Schröder}}]{bopp_sawfish_2024}%
  \BibitemOpen
  \bibfield  {author} {\bibinfo {author} {\bibfnamefont {J.~M.}\ \bibnamefont
  {Bopp}}, \bibinfo {author} {\bibfnamefont {M.}~\bibnamefont {Plock}},
  \bibinfo {author} {\bibfnamefont {T.}~\bibnamefont {Turan}}, \bibinfo
  {author} {\bibfnamefont {G.}~\bibnamefont {Pieplow}}, \bibinfo {author}
  {\bibfnamefont {S.}~\bibnamefont {Burger}}, \ and\ \bibinfo {author}
  {\bibfnamefont {T.}~\bibnamefont {Schröder}},\ }\href
  {https://onlinelibrary.wiley.com/doi/abs/10.1002/adom.202301286} {\bibfield
  {journal} {\bibinfo  {journal} {Adv. Opt. Mater.}\ }\textbf {\bibinfo
  {volume} {12}},\ \bibinfo {pages} {2301286} (\bibinfo {year}
  {2024})}\BibitemShut {NoStop}%
\bibitem [{\citenamefont {Pregnolato}\ \emph {et~al.}(2024)\citenamefont
  {Pregnolato}, \citenamefont {Stucki}, \citenamefont {Bopp}, \citenamefont
  {V.~D.~Hoeven}, \citenamefont {Gokhale}, \citenamefont {Krüger},\ and\
  \citenamefont {Schröder}}]{pregnolato_fabrication_2024}%
  \BibitemOpen
  \bibfield  {author} {\bibinfo {author} {\bibfnamefont {T.}~\bibnamefont
  {Pregnolato}}, \bibinfo {author} {\bibfnamefont {M.~E.}\ \bibnamefont
  {Stucki}}, \bibinfo {author} {\bibfnamefont {J.~M.}\ \bibnamefont {Bopp}},
  \bibinfo {author} {\bibfnamefont {M.~H.}\ \bibnamefont {V.~D.~Hoeven}},
  \bibinfo {author} {\bibfnamefont {A.}~\bibnamefont {Gokhale}}, \bibinfo
  {author} {\bibfnamefont {O.}~\bibnamefont {Krüger}}, \ and\ \bibinfo
  {author} {\bibfnamefont {T.}~\bibnamefont {Schröder}},\ }\href
  {https://pubs.aip.org/app/article/9/3/036105/3269905/Fabrication-of-Sawfish-photonic-crystal-cavities}
  {\bibfield  {journal} {\bibinfo  {journal} {APL Photonics}\ }\textbf
  {\bibinfo {volume} {9}},\ \bibinfo {pages} {036105} (\bibinfo {year}
  {2024})}\BibitemShut {NoStop}%
\bibitem [{\citenamefont {Lee}\ \emph {et~al.}(2018)\citenamefont {Lee},
  \citenamefont {Wells}, \citenamefont {Villa}, \citenamefont {Kalliakos},
  \citenamefont {Stevenson}, \citenamefont {Ellis}, \citenamefont {Farrer},
  \citenamefont {Ritchie}, \citenamefont {Bennett},\ and\ \citenamefont
  {Shields}}]{Lee2018}%
  \BibitemOpen
  \bibfield  {author} {\bibinfo {author} {\bibfnamefont {J.~P.}\ \bibnamefont
  {Lee}}, \bibinfo {author} {\bibfnamefont {L.~M.}\ \bibnamefont {Wells}},
  \bibinfo {author} {\bibfnamefont {B.}~\bibnamefont {Villa}}, \bibinfo
  {author} {\bibfnamefont {S.}~\bibnamefont {Kalliakos}}, \bibinfo {author}
  {\bibfnamefont {R.~M.}\ \bibnamefont {Stevenson}}, \bibinfo {author}
  {\bibfnamefont {D.~J.~P.}\ \bibnamefont {Ellis}}, \bibinfo {author}
  {\bibfnamefont {I.}~\bibnamefont {Farrer}}, \bibinfo {author} {\bibfnamefont
  {D.~A.}\ \bibnamefont {Ritchie}}, \bibinfo {author} {\bibfnamefont {A.~J.}\
  \bibnamefont {Bennett}}, \ and\ \bibinfo {author} {\bibfnamefont {A.~J.}\
  \bibnamefont {Shields}},\ }\href
  {http://dx.doi.org/10.1103/PhysRevX.8.021078} {\bibfield  {journal} {\bibinfo
   {journal} {Phys. Rev. X}\ }\textbf {\bibinfo {volume} {8}} (\bibinfo {year}
  {2018})}\BibitemShut {NoStop}%
\bibitem [{\citenamefont {Bouchard}\ \emph {et~al.}(2022)\citenamefont
  {Bouchard}, \citenamefont {England}, \citenamefont {Bustard}, \citenamefont
  {Heshami},\ and\ \citenamefont {Sussman}}]{Bouchard2022}%
  \BibitemOpen
  \bibfield  {author} {\bibinfo {author} {\bibfnamefont {F.}~\bibnamefont
  {Bouchard}}, \bibinfo {author} {\bibfnamefont {D.}~\bibnamefont {England}},
  \bibinfo {author} {\bibfnamefont {P.~J.}\ \bibnamefont {Bustard}}, \bibinfo
  {author} {\bibfnamefont {K.}~\bibnamefont {Heshami}}, \ and\ \bibinfo
  {author} {\bibfnamefont {B.}~\bibnamefont {Sussman}},\ }\href
  {https://journals.aps.org/prxquantum/abstract/10.1103/PRXQuantum.3.010332}
  {\bibfield  {journal} {\bibinfo  {journal} {Phys. Rev. X Quantum}\ }\textbf
  {\bibinfo {volume} {3}} (\bibinfo {year} {2022})}\BibitemShut {NoStop}%
\bibitem [{\citenamefont {Yu}\ \emph {et~al.}(2025)\citenamefont {Yu},
  \citenamefont {Sciara}, \citenamefont {Chemnitz}, \citenamefont {Montaut},
  \citenamefont {Crockett}, \citenamefont {Fischer}, \citenamefont {Helsten},
  \citenamefont {Wetzel}, \citenamefont {Goebel}, \citenamefont {Kr\"{a}mer},
  \citenamefont {Little}, \citenamefont {Chu}, \citenamefont {Nolte},
  \citenamefont {Wang}, \citenamefont {Azaña}, \citenamefont {Munro},
  \citenamefont {Moss},\ and\ \citenamefont {Morandotti}}]{Yu2025}%
  \BibitemOpen
  \bibfield  {author} {\bibinfo {author} {\bibfnamefont {H.}~\bibnamefont
  {Yu}}, \bibinfo {author} {\bibfnamefont {S.}~\bibnamefont {Sciara}}, \bibinfo
  {author} {\bibfnamefont {M.}~\bibnamefont {Chemnitz}}, \bibinfo {author}
  {\bibfnamefont {N.}~\bibnamefont {Montaut}}, \bibinfo {author} {\bibfnamefont
  {B.}~\bibnamefont {Crockett}}, \bibinfo {author} {\bibfnamefont
  {B.}~\bibnamefont {Fischer}}, \bibinfo {author} {\bibfnamefont
  {R.}~\bibnamefont {Helsten}}, \bibinfo {author} {\bibfnamefont
  {B.}~\bibnamefont {Wetzel}}, \bibinfo {author} {\bibfnamefont {T.~A.}\
  \bibnamefont {Goebel}}, \bibinfo {author} {\bibfnamefont {R.~G.}\
  \bibnamefont {Kr\"{a}mer}}, \bibinfo {author} {\bibfnamefont {B.~E.}\
  \bibnamefont {Little}}, \bibinfo {author} {\bibfnamefont {S.~T.}\
  \bibnamefont {Chu}}, \bibinfo {author} {\bibfnamefont {S.}~\bibnamefont
  {Nolte}}, \bibinfo {author} {\bibfnamefont {Z.}~\bibnamefont {Wang}},
  \bibinfo {author} {\bibfnamefont {J.}~\bibnamefont {Azaña}}, \bibinfo
  {author} {\bibfnamefont {W.~J.}\ \bibnamefont {Munro}}, \bibinfo {author}
  {\bibfnamefont {D.~J.}\ \bibnamefont {Moss}}, \ and\ \bibinfo {author}
  {\bibfnamefont {R.}~\bibnamefont {Morandotti}},\ }\href
  {http://dx.doi.org/10.1038/s41467-024-55345-0} {\bibfield  {journal}
  {\bibinfo  {journal} {Nat. Commun.}\ }\textbf {\bibinfo {volume} {16}}
  (\bibinfo {year} {2025})}\BibitemShut {NoStop}%
\bibitem [{\citenamefont {Strocka}\ \emph
  {et~al.}(2025{\natexlab{a}})\citenamefont {Strocka}, \citenamefont
  {Belhassen}, \citenamefont {Schr\"{o}der},\ and\ \citenamefont
  {Pieplow}}]{strocka_token_2025}%
  \BibitemOpen
  \bibfield  {author} {\bibinfo {author} {\bibfnamefont {Y.}~\bibnamefont
  {Strocka}}, \bibinfo {author} {\bibfnamefont {M.}~\bibnamefont {Belhassen}},
  \bibinfo {author} {\bibfnamefont {T.}~\bibnamefont {Schr\"{o}der}}, \ and\
  \bibinfo {author} {\bibfnamefont {G.}~\bibnamefont {Pieplow}},\ }\href
  {https://arxiv.org/abs/2503.04985} {\bibfield  {journal} {\bibinfo  {journal}
  {arXiv:2503.04985}\ } (\bibinfo {year} {2025}{\natexlab{a}})}\BibitemShut
  {NoStop}%
\bibitem [{\citenamefont {Gisin}\ and\ \citenamefont {Thew}(2007)}]{Gisin2007}%
  \BibitemOpen
  \bibfield  {author} {\bibinfo {author} {\bibfnamefont {N.}~\bibnamefont
  {Gisin}}\ and\ \bibinfo {author} {\bibfnamefont {R.}~\bibnamefont {Thew}},\
  }\href {http://dx.doi.org/10.1038/nphoton.2007.22} {\bibfield  {journal}
  {\bibinfo  {journal} {Nat. Photonics}\ }\textbf {\bibinfo {volume} {1}},\
  \bibinfo {pages} {165–171} (\bibinfo {year} {2007})}\BibitemShut {NoStop}%
\bibitem [{\citenamefont {Chen}(2021)}]{Chen2021}%
  \BibitemOpen
  \bibfield  {author} {\bibinfo {author} {\bibfnamefont {J.}~\bibnamefont
  {Chen}},\ }\href {http://dx.doi.org/10.1088/1742-6596/1865/2/022008}
  {\bibfield  {journal} {\bibinfo  {journal} {J. Phys.: Conf. Ser.}\ }\textbf
  {\bibinfo {volume} {1865}},\ \bibinfo {pages} {022008} (\bibinfo {year}
  {2021})}\BibitemShut {NoStop}%
\bibitem [{\citenamefont {Couteau}\ \emph {et~al.}(2023)\citenamefont
  {Couteau}, \citenamefont {Barz}, \citenamefont {Durt}, \citenamefont
  {Gerrits}, \citenamefont {Huwer}, \citenamefont {Prevedel}, \citenamefont
  {Rarity}, \citenamefont {Shields},\ and\ \citenamefont
  {Weihs}}]{Couteau2023}%
  \BibitemOpen
  \bibfield  {author} {\bibinfo {author} {\bibfnamefont {C.}~\bibnamefont
  {Couteau}}, \bibinfo {author} {\bibfnamefont {S.}~\bibnamefont {Barz}},
  \bibinfo {author} {\bibfnamefont {T.}~\bibnamefont {Durt}}, \bibinfo {author}
  {\bibfnamefont {T.}~\bibnamefont {Gerrits}}, \bibinfo {author} {\bibfnamefont
  {J.}~\bibnamefont {Huwer}}, \bibinfo {author} {\bibfnamefont
  {R.}~\bibnamefont {Prevedel}}, \bibinfo {author} {\bibfnamefont
  {J.}~\bibnamefont {Rarity}}, \bibinfo {author} {\bibfnamefont
  {A.}~\bibnamefont {Shields}}, \ and\ \bibinfo {author} {\bibfnamefont
  {G.}~\bibnamefont {Weihs}},\ }\href
  {http://dx.doi.org/10.1038/s42254-023-00583-2} {\bibfield  {journal}
  {\bibinfo  {journal} {Nat. Rev. Phys.}\ }\textbf {\bibinfo {volume} {5}},\
  \bibinfo {pages} {326–338} (\bibinfo {year} {2023})}\BibitemShut {NoStop}%
\bibitem [{\citenamefont {Gill}\ \emph {et~al.}(2021)\citenamefont {Gill},
  \citenamefont {Kumar}, \citenamefont {Singh}, \citenamefont {Singh},
  \citenamefont {Kaur}, \citenamefont {Usman},\ and\ \citenamefont
  {Buyya}}]{Gill2021}%
  \BibitemOpen
  \bibfield  {author} {\bibinfo {author} {\bibfnamefont {S.~S.}\ \bibnamefont
  {Gill}}, \bibinfo {author} {\bibfnamefont {A.}~\bibnamefont {Kumar}},
  \bibinfo {author} {\bibfnamefont {H.}~\bibnamefont {Singh}}, \bibinfo
  {author} {\bibfnamefont {M.}~\bibnamefont {Singh}}, \bibinfo {author}
  {\bibfnamefont {K.}~\bibnamefont {Kaur}}, \bibinfo {author} {\bibfnamefont
  {M.}~\bibnamefont {Usman}}, \ and\ \bibinfo {author} {\bibfnamefont
  {R.}~\bibnamefont {Buyya}},\ }\href {http://dx.doi.org/10.1002/spe.3039}
  {\bibfield  {journal} {\bibinfo  {journal} {Softw. Pract. Exp.}\ }\textbf
  {\bibinfo {volume} {52}},\ \bibinfo {pages} {66–114} (\bibinfo {year}
  {2021})}\BibitemShut {NoStop}%
\bibitem [{\citenamefont {Sood}\ and\ \citenamefont {Pooja}(2024)}]{Sood2024}%
  \BibitemOpen
  \bibfield  {author} {\bibinfo {author} {\bibfnamefont {S.~K.}\ \bibnamefont
  {Sood}}\ and\ \bibinfo {author} {\bibnamefont {Pooja}},\ }\href
  {http://dx.doi.org/10.1109/TEM.2023.3284689} {\bibfield  {journal} {\bibinfo
  {journal} {IEEE Trans. Eng. Manag.}\ }\textbf {\bibinfo {volume} {71}},\
  \bibinfo {pages} {6662–6676} (\bibinfo {year} {2024})}\BibitemShut
  {NoStop}%
\bibitem [{\citenamefont {Nimbe}, \citenamefont {Weyori},\ and\ \citenamefont
  {Adekoya}(2021)}]{Nimbe2021}%
  \BibitemOpen
  \bibfield  {author} {\bibinfo {author} {\bibfnamefont {P.}~\bibnamefont
  {Nimbe}}, \bibinfo {author} {\bibfnamefont {B.~A.}\ \bibnamefont {Weyori}}, \
  and\ \bibinfo {author} {\bibfnamefont {A.~F.}\ \bibnamefont {Adekoya}},\
  }\href {http://dx.doi.org/10.1007/s11128-021-03021-3} {\bibfield  {journal}
  {\bibinfo  {journal} {Quantum Inf. Process.}\ }\textbf {\bibinfo {volume}
  {20}} (\bibinfo {year} {2021})}\BibitemShut {NoStop}%
\bibitem [{\citenamefont {Degen}, \citenamefont {Reinhard},\ and\ \citenamefont
  {Cappellaro}(2017)}]{Degen2017}%
  \BibitemOpen
  \bibfield  {author} {\bibinfo {author} {\bibfnamefont {C.}~\bibnamefont
  {Degen}}, \bibinfo {author} {\bibfnamefont {F.}~\bibnamefont {Reinhard}}, \
  and\ \bibinfo {author} {\bibfnamefont {P.}~\bibnamefont {Cappellaro}},\
  }\href {http://dx.doi.org/10.1103/RevModPhys.89.035002} {\bibfield  {journal}
  {\bibinfo  {journal} {Rev. Mod. Phys.}\ }\textbf {\bibinfo {volume} {89}}
  (\bibinfo {year} {2017})}\BibitemShut {NoStop}%
\bibitem [{\citenamefont {Zhang}\ and\ \citenamefont
  {Zhuang}(2021)}]{Zhang2021}%
  \BibitemOpen
  \bibfield  {author} {\bibinfo {author} {\bibfnamefont {Z.}~\bibnamefont
  {Zhang}}\ and\ \bibinfo {author} {\bibfnamefont {Q.}~\bibnamefont {Zhuang}},\
  }\href {http://dx.doi.org/10.1088/2058-9565/abd4c3} {\bibfield  {journal}
  {\bibinfo  {journal} {Quantum Sci. Technol.}\ }\textbf {\bibinfo {volume}
  {6}},\ \bibinfo {pages} {043001} (\bibinfo {year} {2021})}\BibitemShut
  {NoStop}%
\bibitem [{\citenamefont {Chang}\ \emph {et~al.}(2023)\citenamefont {Chang},
  \citenamefont {Gao}, \citenamefont {Esmaeil~Zadeh}, \citenamefont
  {Elshaari},\ and\ \citenamefont {Zwiller}}]{Chang2023}%
  \BibitemOpen
  \bibfield  {author} {\bibinfo {author} {\bibfnamefont {J.}~\bibnamefont
  {Chang}}, \bibinfo {author} {\bibfnamefont {J.}~\bibnamefont {Gao}}, \bibinfo
  {author} {\bibfnamefont {I.}~\bibnamefont {Esmaeil~Zadeh}}, \bibinfo {author}
  {\bibfnamefont {A.~W.}\ \bibnamefont {Elshaari}}, \ and\ \bibinfo {author}
  {\bibfnamefont {V.}~\bibnamefont {Zwiller}},\ }\href
  {http://dx.doi.org/10.1515/nanoph-2022-0652} {\bibfield  {journal} {\bibinfo
  {journal} {Nanophotonics}\ }\textbf {\bibinfo {volume} {12}},\ \bibinfo
  {pages} {339–358} (\bibinfo {year} {2023})}\BibitemShut {NoStop}%
\bibitem [{\citenamefont {van Loock}\ \emph {et~al.}(2020)\citenamefont {van
  Loock}, \citenamefont {Alt}, \citenamefont {Becher}, \citenamefont {Benson},
  \citenamefont {Boche}, \citenamefont {Deppe}, \citenamefont {Eschner},
  \citenamefont {H\"{o}fling}, \citenamefont {Meschede}, \citenamefont
  {Michler}, \citenamefont {Schmidt},\ and\ \citenamefont
  {Weinfurter}}]{vanLoock2020}%
  \BibitemOpen
  \bibfield  {author} {\bibinfo {author} {\bibfnamefont {P.}~\bibnamefont {van
  Loock}}, \bibinfo {author} {\bibfnamefont {W.}~\bibnamefont {Alt}}, \bibinfo
  {author} {\bibfnamefont {C.}~\bibnamefont {Becher}}, \bibinfo {author}
  {\bibfnamefont {O.}~\bibnamefont {Benson}}, \bibinfo {author} {\bibfnamefont
  {H.}~\bibnamefont {Boche}}, \bibinfo {author} {\bibfnamefont
  {C.}~\bibnamefont {Deppe}}, \bibinfo {author} {\bibfnamefont
  {J.}~\bibnamefont {Eschner}}, \bibinfo {author} {\bibfnamefont
  {S.}~\bibnamefont {H\"{o}fling}}, \bibinfo {author} {\bibfnamefont
  {D.}~\bibnamefont {Meschede}}, \bibinfo {author} {\bibfnamefont
  {P.}~\bibnamefont {Michler}}, \bibinfo {author} {\bibfnamefont
  {F.}~\bibnamefont {Schmidt}}, \ and\ \bibinfo {author} {\bibfnamefont
  {H.}~\bibnamefont {Weinfurter}},\ }\href
  {http://dx.doi.org/10.1002/qute.201900141} {\bibfield  {journal} {\bibinfo
  {journal} {Adv. Quantum Technol.}\ }\textbf {\bibinfo {volume} {3}} (\bibinfo
  {year} {2020})}\BibitemShut {NoStop}%
\bibitem [{\citenamefont {Yan}\ \emph {et~al.}(2021)\citenamefont {Yan},
  \citenamefont {Zhou}, \citenamefont {Zhong},\ and\ \citenamefont
  {Sheng}}]{Yan2021}%
  \BibitemOpen
  \bibfield  {author} {\bibinfo {author} {\bibfnamefont {P.-S.}\ \bibnamefont
  {Yan}}, \bibinfo {author} {\bibfnamefont {L.}~\bibnamefont {Zhou}}, \bibinfo
  {author} {\bibfnamefont {W.}~\bibnamefont {Zhong}}, \ and\ \bibinfo {author}
  {\bibfnamefont {Y.-B.}\ \bibnamefont {Sheng}},\ }\href
  {http://dx.doi.org/10.1209/0295-5075/ac37d0} {\bibfield  {journal} {\bibinfo
  {journal} {EPL}\ }\textbf {\bibinfo {volume} {136}},\ \bibinfo {pages}
  {14001} (\bibinfo {year} {2021})}\BibitemShut {NoStop}%
\bibitem [{\citenamefont {Azuma}\ \emph {et~al.}(2023)\citenamefont {Azuma},
  \citenamefont {Economou}, \citenamefont {Elkouss}, \citenamefont {Hilaire},
  \citenamefont {Jiang}, \citenamefont {Lo},\ and\ \citenamefont
  {Tzitrin}}]{Azuma2023}%
  \BibitemOpen
  \bibfield  {author} {\bibinfo {author} {\bibfnamefont {K.}~\bibnamefont
  {Azuma}}, \bibinfo {author} {\bibfnamefont {S.~E.}\ \bibnamefont {Economou}},
  \bibinfo {author} {\bibfnamefont {D.}~\bibnamefont {Elkouss}}, \bibinfo
  {author} {\bibfnamefont {P.}~\bibnamefont {Hilaire}}, \bibinfo {author}
  {\bibfnamefont {L.}~\bibnamefont {Jiang}}, \bibinfo {author} {\bibfnamefont
  {H.-K.}\ \bibnamefont {Lo}}, \ and\ \bibinfo {author} {\bibfnamefont
  {I.}~\bibnamefont {Tzitrin}},\ }\href
  {http://dx.doi.org/10.1103/RevModPhys.95.045006} {\bibfield  {journal}
  {\bibinfo  {journal} {Rev. Mod. Phys.}\ }\textbf {\bibinfo {volume} {95}}
  (\bibinfo {year} {2023})}\BibitemShut {NoStop}%
\bibitem [{\citenamefont {Portmann}\ and\ \citenamefont
  {Renner}(2022)}]{Portmann2022}%
  \BibitemOpen
  \bibfield  {author} {\bibinfo {author} {\bibfnamefont {C.}~\bibnamefont
  {Portmann}}\ and\ \bibinfo {author} {\bibfnamefont {R.}~\bibnamefont
  {Renner}},\ }\href {http://dx.doi.org/10.1103/RevModPhys.94.025008}
  {\bibfield  {journal} {\bibinfo  {journal} {Rev. Mod. Phys.}\ }\textbf
  {\bibinfo {volume} {94}} (\bibinfo {year} {2022})}\BibitemShut {NoStop}%
\bibitem [{\citenamefont {Dutta}\ and\ \citenamefont
  {Pathak}(2022)}]{Dutta2022}%
  \BibitemOpen
  \bibfield  {author} {\bibinfo {author} {\bibfnamefont {A.}~\bibnamefont
  {Dutta}}\ and\ \bibinfo {author} {\bibfnamefont {A.}~\bibnamefont {Pathak}},\
  }\href {http://dx.doi.org/10.1007/s11128-022-03717-0} {\bibfield  {journal}
  {\bibinfo  {journal} {Quantum Inf. Process.}\ }\textbf {\bibinfo {volume}
  {21}} (\bibinfo {year} {2022})}\BibitemShut {NoStop}%
\bibitem [{\citenamefont {Kyaw}\ \emph {et~al.}(2015)\citenamefont {Kyaw},
  \citenamefont {Felicetti}, \citenamefont {Romero}, \citenamefont {Solano},\
  and\ \citenamefont {Kwek}}]{Kyaw2015}%
  \BibitemOpen
  \bibfield  {author} {\bibinfo {author} {\bibfnamefont {T.~H.}\ \bibnamefont
  {Kyaw}}, \bibinfo {author} {\bibfnamefont {S.}~\bibnamefont {Felicetti}},
  \bibinfo {author} {\bibfnamefont {G.}~\bibnamefont {Romero}}, \bibinfo
  {author} {\bibfnamefont {E.}~\bibnamefont {Solano}}, \ and\ \bibinfo {author}
  {\bibfnamefont {L.-C.}\ \bibnamefont {Kwek}},\ }\href
  {http://dx.doi.org/10.1038/srep08621} {\bibfield  {journal} {\bibinfo
  {journal} {Sci. Rep.}\ }\textbf {\bibinfo {volume} {5}} (\bibinfo {year}
  {2015})}\BibitemShut {NoStop}%
\bibitem [{\citenamefont {Jaques}\ and\ \citenamefont
  {Rattew}(2025)}]{qram2023}%
  \BibitemOpen
  \bibfield  {author} {\bibinfo {author} {\bibfnamefont {S.}~\bibnamefont
  {Jaques}}\ and\ \bibinfo {author} {\bibfnamefont {A.~G.}\ \bibnamefont
  {Rattew}},\ }\href {http://dx.doi.org/10.22331/q-2025-12-02-1922} {\bibfield
  {journal} {\bibinfo  {journal} {Quantum}\ }\textbf {\bibinfo {volume} {9}},\
  \bibinfo {pages} {1922} (\bibinfo {year} {2025})}\BibitemShut {NoStop}%
\bibitem [{\citenamefont {Zaiser}\ \emph {et~al.}(2016)\citenamefont {Zaiser},
  \citenamefont {Rendler}, \citenamefont {Jakobi}, \citenamefont {Wolf},
  \citenamefont {Lee}, \citenamefont {Wagner}, \citenamefont {Bergholm},
  \citenamefont {Schulte-Herbr\"{u}ggen}, \citenamefont {Neumann},\ and\
  \citenamefont {Wrachtrup}}]{Zaiser2016}%
  \BibitemOpen
  \bibfield  {author} {\bibinfo {author} {\bibfnamefont {S.}~\bibnamefont
  {Zaiser}}, \bibinfo {author} {\bibfnamefont {T.}~\bibnamefont {Rendler}},
  \bibinfo {author} {\bibfnamefont {I.}~\bibnamefont {Jakobi}}, \bibinfo
  {author} {\bibfnamefont {T.}~\bibnamefont {Wolf}}, \bibinfo {author}
  {\bibfnamefont {S.-Y.}\ \bibnamefont {Lee}}, \bibinfo {author} {\bibfnamefont
  {S.}~\bibnamefont {Wagner}}, \bibinfo {author} {\bibfnamefont
  {V.}~\bibnamefont {Bergholm}}, \bibinfo {author} {\bibfnamefont
  {T.}~\bibnamefont {Schulte-Herbr\"{u}ggen}}, \bibinfo {author} {\bibfnamefont
  {P.}~\bibnamefont {Neumann}}, \ and\ \bibinfo {author} {\bibfnamefont
  {J.}~\bibnamefont {Wrachtrup}},\ }\href
  {http://dx.doi.org/10.1038/ncomms12279} {\bibfield  {journal} {\bibinfo
  {journal} {Nat. Commun.}\ }\textbf {\bibinfo {volume} {7}} (\bibinfo {year}
  {2016})}\BibitemShut {NoStop}%
\bibitem [{\citenamefont {Tanji}\ \emph {et~al.}(2009)\citenamefont {Tanji},
  \citenamefont {Simon}, \citenamefont {Ghosh}, \citenamefont {Bloom},\ and\
  \citenamefont {Vuletić}}]{Tanji2009}%
  \BibitemOpen
  \bibfield  {author} {\bibinfo {author} {\bibfnamefont {H.}~\bibnamefont
  {Tanji}}, \bibinfo {author} {\bibfnamefont {J.}~\bibnamefont {Simon}},
  \bibinfo {author} {\bibfnamefont {S.}~\bibnamefont {Ghosh}}, \bibinfo
  {author} {\bibfnamefont {B.}~\bibnamefont {Bloom}}, \ and\ \bibinfo {author}
  {\bibfnamefont {V.}~\bibnamefont {Vuletić}},\ }\href
  {http://dx.doi.org/10.1088/0031-8949/2009/T135/014010} {\bibfield  {journal}
  {\bibinfo  {journal} {Phys. Scr.}\ }\textbf {\bibinfo {volume} {T135}},\
  \bibinfo {pages} {014010} (\bibinfo {year} {2009})}\BibitemShut {NoStop}%
\bibitem [{\citenamefont {Vernaz-Gris}\ \emph {et~al.}(2018)\citenamefont
  {Vernaz-Gris}, \citenamefont {Huang}, \citenamefont {Cao}, \citenamefont
  {Sheremet},\ and\ \citenamefont {Laurat}}]{VernazGris2018}%
  \BibitemOpen
  \bibfield  {author} {\bibinfo {author} {\bibfnamefont {P.}~\bibnamefont
  {Vernaz-Gris}}, \bibinfo {author} {\bibfnamefont {K.}~\bibnamefont {Huang}},
  \bibinfo {author} {\bibfnamefont {M.}~\bibnamefont {Cao}}, \bibinfo {author}
  {\bibfnamefont {A.~S.}\ \bibnamefont {Sheremet}}, \ and\ \bibinfo {author}
  {\bibfnamefont {J.}~\bibnamefont {Laurat}},\ }\href
  {http://dx.doi.org/10.1038/s41467-017-02775-8} {\bibfield  {journal}
  {\bibinfo  {journal} {Nat. Commun.}\ }\textbf {\bibinfo {volume} {9}}
  (\bibinfo {year} {2018})}\BibitemShut {NoStop}%
\bibitem [{\citenamefont {Lago-Rivera}\ \emph {et~al.}(2021)\citenamefont
  {Lago-Rivera}, \citenamefont {Grandi}, \citenamefont {Rakonjac},
  \citenamefont {Seri},\ and\ \citenamefont {de~Riedmatten}}]{LagoRivera2021}%
  \BibitemOpen
  \bibfield  {author} {\bibinfo {author} {\bibfnamefont {D.}~\bibnamefont
  {Lago-Rivera}}, \bibinfo {author} {\bibfnamefont {S.}~\bibnamefont {Grandi}},
  \bibinfo {author} {\bibfnamefont {J.~V.}\ \bibnamefont {Rakonjac}}, \bibinfo
  {author} {\bibfnamefont {A.}~\bibnamefont {Seri}}, \ and\ \bibinfo {author}
  {\bibfnamefont {H.}~\bibnamefont {de~Riedmatten}},\ }\href
  {http://dx.doi.org/10.1038/s41586-021-03481-8} {\bibfield  {journal}
  {\bibinfo  {journal} {Nature}\ }\textbf {\bibinfo {volume} {594}},\ \bibinfo
  {pages} {37–40} (\bibinfo {year} {2021})}\BibitemShut {NoStop}%
\bibitem [{\citenamefont {Bradley}\ \emph {et~al.}(2022)\citenamefont
  {Bradley}, \citenamefont {de~Bone}, \citenamefont {M\"{o}ller}, \citenamefont
  {Baier}, \citenamefont {Degen}, \citenamefont {Loenen}, \citenamefont
  {Bartling}, \citenamefont {Markham}, \citenamefont {Twitchen}, \citenamefont
  {Hanson}, \citenamefont {Elkouss},\ and\ \citenamefont
  {Taminiau}}]{Bradley2022}%
  \BibitemOpen
  \bibfield  {author} {\bibinfo {author} {\bibfnamefont {C.~E.}\ \bibnamefont
  {Bradley}}, \bibinfo {author} {\bibfnamefont {S.~W.}\ \bibnamefont
  {de~Bone}}, \bibinfo {author} {\bibfnamefont {P.~F.~W.}\ \bibnamefont
  {M\"{o}ller}}, \bibinfo {author} {\bibfnamefont {S.}~\bibnamefont {Baier}},
  \bibinfo {author} {\bibfnamefont {M.~J.}\ \bibnamefont {Degen}}, \bibinfo
  {author} {\bibfnamefont {S.~J.~H.}\ \bibnamefont {Loenen}}, \bibinfo {author}
  {\bibfnamefont {H.~P.}\ \bibnamefont {Bartling}}, \bibinfo {author}
  {\bibfnamefont {M.}~\bibnamefont {Markham}}, \bibinfo {author} {\bibfnamefont
  {D.~J.}\ \bibnamefont {Twitchen}}, \bibinfo {author} {\bibfnamefont
  {R.}~\bibnamefont {Hanson}}, \bibinfo {author} {\bibfnamefont
  {D.}~\bibnamefont {Elkouss}}, \ and\ \bibinfo {author} {\bibfnamefont
  {T.~H.}\ \bibnamefont {Taminiau}},\ }\href
  {http://dx.doi.org/10.1038/s41534-022-00637-w} {\bibfield  {journal}
  {\bibinfo  {journal} {npj Quantum Inf.}\ }\textbf {\bibinfo {volume} {8}}
  (\bibinfo {year} {2022})}\BibitemShut {NoStop}%
\bibitem [{\citenamefont {Bhaskar}\ \emph
  {et~al.}(2020{\natexlab{a}})\citenamefont {Bhaskar}, \citenamefont
  {Riedinger}, \citenamefont {Machielse}, \citenamefont {Levonian},
  \citenamefont {Nguyen}, \citenamefont {Knall}, \citenamefont {Park},
  \citenamefont {Englund}, \citenamefont {Lončar}, \citenamefont {Sukachev},\
  and\ \citenamefont {Lukin}}]{Bhaskar2020}%
  \BibitemOpen
  \bibfield  {author} {\bibinfo {author} {\bibfnamefont {M.~K.}\ \bibnamefont
  {Bhaskar}}, \bibinfo {author} {\bibfnamefont {R.}~\bibnamefont {Riedinger}},
  \bibinfo {author} {\bibfnamefont {B.}~\bibnamefont {Machielse}}, \bibinfo
  {author} {\bibfnamefont {D.~S.}\ \bibnamefont {Levonian}}, \bibinfo {author}
  {\bibfnamefont {C.~T.}\ \bibnamefont {Nguyen}}, \bibinfo {author}
  {\bibfnamefont {E.~N.}\ \bibnamefont {Knall}}, \bibinfo {author}
  {\bibfnamefont {H.}~\bibnamefont {Park}}, \bibinfo {author} {\bibfnamefont
  {D.}~\bibnamefont {Englund}}, \bibinfo {author} {\bibfnamefont
  {M.}~\bibnamefont {Lončar}}, \bibinfo {author} {\bibfnamefont {D.~D.}\
  \bibnamefont {Sukachev}}, \ and\ \bibinfo {author} {\bibfnamefont {M.~D.}\
  \bibnamefont {Lukin}},\ }\href {\doibase 10.1038/s41586-020-2103-5}
  {\bibfield  {journal} {\bibinfo  {journal} {Nature}\ }\textbf {\bibinfo
  {volume} {580}},\ \bibinfo {pages} {60–64} (\bibinfo {year}
  {2020}{\natexlab{a}})}\BibitemShut {NoStop}%
\bibitem [{\citenamefont {Bonarota}\ \emph {et~al.}(2010)\citenamefont
  {Bonarota}, \citenamefont {Ruggiero}, \citenamefont {Gouët},\ and\
  \citenamefont {Chanelière}}]{Bonarota2010}%
  \BibitemOpen
  \bibfield  {author} {\bibinfo {author} {\bibfnamefont {M.}~\bibnamefont
  {Bonarota}}, \bibinfo {author} {\bibfnamefont {J.}~\bibnamefont {Ruggiero}},
  \bibinfo {author} {\bibfnamefont {J.~L.~L.}\ \bibnamefont {Gouët}}, \ and\
  \bibinfo {author} {\bibfnamefont {T.}~\bibnamefont {Chanelière}},\ }\href
  {http://dx.doi.org/10.1103/PhysRevA.81.033803} {\bibfield  {journal}
  {\bibinfo  {journal} {Phys. Rev. A}\ }\textbf {\bibinfo {volume} {81}}
  (\bibinfo {year} {2010})}\BibitemShut {NoStop}%
\bibitem [{\citenamefont {Wendin}(2017)}]{Wendin2017}%
  \BibitemOpen
  \bibfield  {author} {\bibinfo {author} {\bibfnamefont {G.}~\bibnamefont
  {Wendin}},\ }\href {http://dx.doi.org/10.1088/1361-6633/aa7e1a} {\bibfield
  {journal} {\bibinfo  {journal} {Rep. Prog. Phys.}\ }\textbf {\bibinfo
  {volume} {80}},\ \bibinfo {pages} {106001} (\bibinfo {year}
  {2017})}\BibitemShut {NoStop}%
\bibitem [{\citenamefont {Bao}\ \emph {et~al.}(2021)\citenamefont {Bao},
  \citenamefont {Wang}, \citenamefont {Wu}, \citenamefont {Li}, \citenamefont
  {Ma}, \citenamefont {Song}, \citenamefont {Zhang},\ and\ \citenamefont
  {Duan}}]{Bao2021}%
  \BibitemOpen
  \bibfield  {author} {\bibinfo {author} {\bibfnamefont {Z.}~\bibnamefont
  {Bao}}, \bibinfo {author} {\bibfnamefont {Z.}~\bibnamefont {Wang}}, \bibinfo
  {author} {\bibfnamefont {Y.}~\bibnamefont {Wu}}, \bibinfo {author}
  {\bibfnamefont {Y.}~\bibnamefont {Li}}, \bibinfo {author} {\bibfnamefont
  {C.}~\bibnamefont {Ma}}, \bibinfo {author} {\bibfnamefont {Y.}~\bibnamefont
  {Song}}, \bibinfo {author} {\bibfnamefont {H.}~\bibnamefont {Zhang}}, \ and\
  \bibinfo {author} {\bibfnamefont {L.}~\bibnamefont {Duan}},\ }\href
  {http://dx.doi.org/10.1103/PhysRevLett.127.010503} {\bibfield  {journal}
  {\bibinfo  {journal} {Phys. Rev. Lett.}\ }\textbf {\bibinfo {volume} {127}}
  (\bibinfo {year} {2021})}\BibitemShut {NoStop}%
\bibitem [{\citenamefont {Matanin}\ \emph {et~al.}(2023)\citenamefont
  {Matanin}, \citenamefont {Gerasimov}, \citenamefont {Moiseev}, \citenamefont
  {Smirnov}, \citenamefont {Ivanov}, \citenamefont {Malevannaya}, \citenamefont
  {Polozov}, \citenamefont {Zikiy}, \citenamefont {Samoilov}, \citenamefont
  {Rodionov},\ and\ \citenamefont {Moiseev}}]{Matanin2023}%
  \BibitemOpen
  \bibfield  {author} {\bibinfo {author} {\bibfnamefont {A.~R.}\ \bibnamefont
  {Matanin}}, \bibinfo {author} {\bibfnamefont {K.~I.}\ \bibnamefont
  {Gerasimov}}, \bibinfo {author} {\bibfnamefont {E.~S.}\ \bibnamefont
  {Moiseev}}, \bibinfo {author} {\bibfnamefont {N.~S.}\ \bibnamefont
  {Smirnov}}, \bibinfo {author} {\bibfnamefont {A.~I.}\ \bibnamefont {Ivanov}},
  \bibinfo {author} {\bibfnamefont {E.~I.}\ \bibnamefont {Malevannaya}},
  \bibinfo {author} {\bibfnamefont {V.~I.}\ \bibnamefont {Polozov}}, \bibinfo
  {author} {\bibfnamefont {E.~V.}\ \bibnamefont {Zikiy}}, \bibinfo {author}
  {\bibfnamefont {A.~A.}\ \bibnamefont {Samoilov}}, \bibinfo {author}
  {\bibfnamefont {I.~A.}\ \bibnamefont {Rodionov}}, \ and\ \bibinfo {author}
  {\bibfnamefont {S.~A.}\ \bibnamefont {Moiseev}},\ }\href
  {http://dx.doi.org/10.1103/PhysRevApplied.19.034011} {\bibfield  {journal}
  {\bibinfo  {journal} {Phys. Rev. Appl.}\ }\textbf {\bibinfo {volume} {19}}
  (\bibinfo {year} {2023})}\BibitemShut {NoStop}%
\bibitem [{\citenamefont {Ganjam}\ \emph {et~al.}(2024)\citenamefont {Ganjam},
  \citenamefont {Wang}, \citenamefont {Lu}, \citenamefont {Banerjee},
  \citenamefont {Lei}, \citenamefont {Krayzman}, \citenamefont {Kisslinger},
  \citenamefont {Zhou}, \citenamefont {Li}, \citenamefont {Jia}, \citenamefont
  {Liu}, \citenamefont {Frunzio},\ and\ \citenamefont
  {Schoelkopf}}]{Ganjam2024}%
  \BibitemOpen
  \bibfield  {author} {\bibinfo {author} {\bibfnamefont {S.}~\bibnamefont
  {Ganjam}}, \bibinfo {author} {\bibfnamefont {Y.}~\bibnamefont {Wang}},
  \bibinfo {author} {\bibfnamefont {Y.}~\bibnamefont {Lu}}, \bibinfo {author}
  {\bibfnamefont {A.}~\bibnamefont {Banerjee}}, \bibinfo {author}
  {\bibfnamefont {C.~U.}\ \bibnamefont {Lei}}, \bibinfo {author} {\bibfnamefont
  {L.}~\bibnamefont {Krayzman}}, \bibinfo {author} {\bibfnamefont
  {K.}~\bibnamefont {Kisslinger}}, \bibinfo {author} {\bibfnamefont
  {C.}~\bibnamefont {Zhou}}, \bibinfo {author} {\bibfnamefont {R.}~\bibnamefont
  {Li}}, \bibinfo {author} {\bibfnamefont {Y.}~\bibnamefont {Jia}}, \bibinfo
  {author} {\bibfnamefont {M.}~\bibnamefont {Liu}}, \bibinfo {author}
  {\bibfnamefont {L.}~\bibnamefont {Frunzio}}, \ and\ \bibinfo {author}
  {\bibfnamefont {R.~J.}\ \bibnamefont {Schoelkopf}},\ }\href
  {https://www.nature.com/articles/s41467-024-47857-6} {\bibfield  {journal}
  {\bibinfo  {journal} {Nat. Commun.}\ }\textbf {\bibinfo {volume} {15}},\
  \bibinfo {pages} {3687} (\bibinfo {year} {2024})}\BibitemShut {NoStop}%
\bibitem [{\citenamefont {Milul}\ \emph {et~al.}(2023)\citenamefont {Milul},
  \citenamefont {Guttel}, \citenamefont {Goldblatt}, \citenamefont {Hazanov},
  \citenamefont {Joshi}, \citenamefont {Chausovsky}, \citenamefont {Kahn},
  \citenamefont {undefinedifty\"{u}rek}, \citenamefont {Lafont},\ and\
  \citenamefont {Rosenblum}}]{Milul2023}%
  \BibitemOpen
  \bibfield  {author} {\bibinfo {author} {\bibfnamefont {O.}~\bibnamefont
  {Milul}}, \bibinfo {author} {\bibfnamefont {B.}~\bibnamefont {Guttel}},
  \bibinfo {author} {\bibfnamefont {U.}~\bibnamefont {Goldblatt}}, \bibinfo
  {author} {\bibfnamefont {S.}~\bibnamefont {Hazanov}}, \bibinfo {author}
  {\bibfnamefont {L.~M.}\ \bibnamefont {Joshi}}, \bibinfo {author}
  {\bibfnamefont {D.}~\bibnamefont {Chausovsky}}, \bibinfo {author}
  {\bibfnamefont {N.}~\bibnamefont {Kahn}}, \bibinfo {author} {\bibfnamefont
  {E.}~\bibnamefont {undefinedifty\"{u}rek}}, \bibinfo {author} {\bibfnamefont
  {F.}~\bibnamefont {Lafont}}, \ and\ \bibinfo {author} {\bibfnamefont
  {S.}~\bibnamefont {Rosenblum}},\ }\href
  {http://dx.doi.org/10.1103/PRXQuantum.4.030336} {\bibfield  {journal}
  {\bibinfo  {journal} {PRX Quantum}\ }\textbf {\bibinfo {volume} {4}}
  (\bibinfo {year} {2023})}\BibitemShut {NoStop}%
\bibitem [{\citenamefont {Wang}\ \emph {et~al.}(2022)\citenamefont {Wang},
  \citenamefont {Gonin}, \citenamefont {Grassellino}, \citenamefont {Kazakov},
  \citenamefont {Romanenko}, \citenamefont {Yakovlev},\ and\ \citenamefont
  {Zorzetti}}]{Wang2022}%
  \BibitemOpen
  \bibfield  {author} {\bibinfo {author} {\bibfnamefont {C.}~\bibnamefont
  {Wang}}, \bibinfo {author} {\bibfnamefont {I.}~\bibnamefont {Gonin}},
  \bibinfo {author} {\bibfnamefont {A.}~\bibnamefont {Grassellino}}, \bibinfo
  {author} {\bibfnamefont {S.}~\bibnamefont {Kazakov}}, \bibinfo {author}
  {\bibfnamefont {A.}~\bibnamefont {Romanenko}}, \bibinfo {author}
  {\bibfnamefont {V.~P.}\ \bibnamefont {Yakovlev}}, \ and\ \bibinfo {author}
  {\bibfnamefont {S.}~\bibnamefont {Zorzetti}},\ }\href
  {http://dx.doi.org/10.1038/s41534-022-00664-7} {\bibfield  {journal}
  {\bibinfo  {journal} {npj Quantum Inf.}\ }\textbf {\bibinfo {volume} {8}}
  (\bibinfo {year} {2022})}\BibitemShut {NoStop}%
\bibitem [{\citenamefont {Magnard}\ \emph {et~al.}(2020)\citenamefont
  {Magnard}, \citenamefont {Storz}, \citenamefont {Kurpiers}, \citenamefont
  {Schär}, \citenamefont {Marxer}, \citenamefont {Lütolf}, \citenamefont
  {Walter}, \citenamefont {Besse}, \citenamefont {Gabureac}, \citenamefont
  {Reuer}, \citenamefont {Akin}, \citenamefont {Royer}, \citenamefont {Blais},\
  and\ \citenamefont {Wallraff}}]{magnard_microwave_2020}%
  \BibitemOpen
  \bibfield  {author} {\bibinfo {author} {\bibfnamefont {P.}~\bibnamefont
  {Magnard}}, \bibinfo {author} {\bibfnamefont {S.}~\bibnamefont {Storz}},
  \bibinfo {author} {\bibfnamefont {P.}~\bibnamefont {Kurpiers}}, \bibinfo
  {author} {\bibfnamefont {J.}~\bibnamefont {Schär}}, \bibinfo {author}
  {\bibfnamefont {F.}~\bibnamefont {Marxer}}, \bibinfo {author} {\bibfnamefont
  {J.}~\bibnamefont {Lütolf}}, \bibinfo {author} {\bibfnamefont
  {T.}~\bibnamefont {Walter}}, \bibinfo {author} {\bibfnamefont {J.-C.}\
  \bibnamefont {Besse}}, \bibinfo {author} {\bibfnamefont {M.}~\bibnamefont
  {Gabureac}}, \bibinfo {author} {\bibfnamefont {K.}~\bibnamefont {Reuer}},
  \bibinfo {author} {\bibfnamefont {A.}~\bibnamefont {Akin}}, \bibinfo {author}
  {\bibfnamefont {B.}~\bibnamefont {Royer}}, \bibinfo {author} {\bibfnamefont
  {A.}~\bibnamefont {Blais}}, \ and\ \bibinfo {author} {\bibfnamefont
  {A.}~\bibnamefont {Wallraff}},\ }\href {\doibase
  10.1103/PhysRevLett.125.260502} {\bibfield  {journal} {\bibinfo  {journal}
  {Phys. Rev. Lett.}\ }\textbf {\bibinfo {volume} {125}},\ \bibinfo {pages}
  {260502} (\bibinfo {year} {2020})}\BibitemShut {NoStop}%
\bibitem [{\citenamefont {Arnold}\ \emph {et~al.}(2025)\citenamefont {Arnold},
  \citenamefont {Werner}, \citenamefont {Sahu}, \citenamefont {Kapoor},
  \citenamefont {Qiu},\ and\ \citenamefont {Fink}}]{Arnold2025}%
  \BibitemOpen
  \bibfield  {author} {\bibinfo {author} {\bibfnamefont {G.}~\bibnamefont
  {Arnold}}, \bibinfo {author} {\bibfnamefont {T.}~\bibnamefont {Werner}},
  \bibinfo {author} {\bibfnamefont {R.}~\bibnamefont {Sahu}}, \bibinfo {author}
  {\bibfnamefont {L.~N.}\ \bibnamefont {Kapoor}}, \bibinfo {author}
  {\bibfnamefont {L.}~\bibnamefont {Qiu}}, \ and\ \bibinfo {author}
  {\bibfnamefont {J.~M.}\ \bibnamefont {Fink}},\ }\href
  {http://dx.doi.org/10.1038/s41567-024-02741-4} {\bibfield  {journal}
  {\bibinfo  {journal} {Nat. Physics}\ }\textbf {\bibinfo {volume} {21}},\
  \bibinfo {pages} {393–400} (\bibinfo {year} {2025})}\BibitemShut {NoStop}%
\bibitem [{\citenamefont {Thomas}\ \emph {et~al.}(2024)\citenamefont {Thomas},
  \citenamefont {Wagner}, \citenamefont {Joos}, \citenamefont {Sittig},
  \citenamefont {Nawrath}, \citenamefont {Burdekin}, \citenamefont
  {de~Buy~Wenniger}, \citenamefont {Rasiah}, \citenamefont {Huber-Loyola},
  \citenamefont {Sagona-Stophel}, \citenamefont {H\"{o}fling}, \citenamefont
  {Jetter}, \citenamefont {Michler}, \citenamefont {Walmsley}, \citenamefont
  {Portalupi},\ and\ \citenamefont {Ledingham}}]{Thomas2024}%
  \BibitemOpen
  \bibfield  {author} {\bibinfo {author} {\bibfnamefont {S.~E.}\ \bibnamefont
  {Thomas}}, \bibinfo {author} {\bibfnamefont {L.}~\bibnamefont {Wagner}},
  \bibinfo {author} {\bibfnamefont {R.}~\bibnamefont {Joos}}, \bibinfo {author}
  {\bibfnamefont {R.}~\bibnamefont {Sittig}}, \bibinfo {author} {\bibfnamefont
  {C.}~\bibnamefont {Nawrath}}, \bibinfo {author} {\bibfnamefont
  {P.}~\bibnamefont {Burdekin}}, \bibinfo {author} {\bibfnamefont {I.~M.}\
  \bibnamefont {de~Buy~Wenniger}}, \bibinfo {author} {\bibfnamefont {M.~J.}\
  \bibnamefont {Rasiah}}, \bibinfo {author} {\bibfnamefont {T.}~\bibnamefont
  {Huber-Loyola}}, \bibinfo {author} {\bibfnamefont {S.}~\bibnamefont
  {Sagona-Stophel}}, \bibinfo {author} {\bibfnamefont {S.}~\bibnamefont
  {H\"{o}fling}}, \bibinfo {author} {\bibfnamefont {M.}~\bibnamefont {Jetter}},
  \bibinfo {author} {\bibfnamefont {P.}~\bibnamefont {Michler}}, \bibinfo
  {author} {\bibfnamefont {I.~A.}\ \bibnamefont {Walmsley}}, \bibinfo {author}
  {\bibfnamefont {S.~L.}\ \bibnamefont {Portalupi}}, \ and\ \bibinfo {author}
  {\bibfnamefont {P.~M.}\ \bibnamefont {Ledingham}},\ }\href
  {http://dx.doi.org/10.1126/sciadv.adi7346} {\bibfield  {journal} {\bibinfo
  {journal} {Science Advances}\ }\textbf {\bibinfo {volume} {10}} (\bibinfo
  {year} {2024})}\BibitemShut {NoStop}%
\bibitem [{\citenamefont {Kaczmarek}(2017)}]{kaczmarek2017a}%
  \BibitemOpen
  \bibfield  {author} {\bibinfo {author} {\bibfnamefont {K.}~\bibnamefont
  {Kaczmarek}},\ }\emph {\bibinfo {title} {ORCA - towards an integrated
  noise-free quantum memory}},\ \href
  {https://ora.ox.ac.uk/objects/uuid:1b4c7463-6181-4689-87d8-5988d4c5bc48}
  {Ph.D. thesis},\ \bibinfo  {school} {University of Oxford} (\bibinfo {year}
  {2017})\BibitemShut {NoStop}%
\bibitem [{\citenamefont {Reimer}\ and\ \citenamefont
  {Cher}(2019)}]{Reimer2019}%
  \BibitemOpen
  \bibfield  {author} {\bibinfo {author} {\bibfnamefont {M.~E.}\ \bibnamefont
  {Reimer}}\ and\ \bibinfo {author} {\bibfnamefont {C.}~\bibnamefont {Cher}},\
  }\href {\doibase 10.1038/s41566-019-0544-x} {\bibfield  {journal} {\bibinfo
  {journal} {Nat. Photonics}\ }\textbf {\bibinfo {volume} {13}},\ \bibinfo
  {pages} {734–736} (\bibinfo {year} {2019})}\BibitemShut {NoStop}%
\bibitem [{\citenamefont {Meng}\ \emph {et~al.}(2024)\citenamefont {Meng},
  \citenamefont {Liu}, \citenamefont {Jin}, \citenamefont {Zhou}, \citenamefont
  {Li},\ and\ \citenamefont {Guo}}]{Meng2024}%
  \BibitemOpen
  \bibfield  {author} {\bibinfo {author} {\bibfnamefont {R.-R.}\ \bibnamefont
  {Meng}}, \bibinfo {author} {\bibfnamefont {X.}~\bibnamefont {Liu}}, \bibinfo
  {author} {\bibfnamefont {M.}~\bibnamefont {Jin}}, \bibinfo {author}
  {\bibfnamefont {Z.-Q.}\ \bibnamefont {Zhou}}, \bibinfo {author}
  {\bibfnamefont {C.-F.}\ \bibnamefont {Li}}, \ and\ \bibinfo {author}
  {\bibfnamefont {G.-C.}\ \bibnamefont {Guo}},\ }\href
  {http://dx.doi.org/10.1016/j.chip.2023.100081} {\bibfield  {journal}
  {\bibinfo  {journal} {Chip}\ }\textbf {\bibinfo {volume} {3}},\ \bibinfo
  {pages} {100081} (\bibinfo {year} {2024})}\BibitemShut {NoStop}%
\bibitem [{\citenamefont {Kaneda}\ \emph {et~al.}(2016)\citenamefont {Kaneda},
  \citenamefont {Garay-Palmett}, \citenamefont {U’Ren},\ and\ \citenamefont
  {Kwiat}}]{Kaneda2016}%
  \BibitemOpen
  \bibfield  {author} {\bibinfo {author} {\bibfnamefont {F.}~\bibnamefont
  {Kaneda}}, \bibinfo {author} {\bibfnamefont {K.}~\bibnamefont
  {Garay-Palmett}}, \bibinfo {author} {\bibfnamefont {A.~B.}\ \bibnamefont
  {U’Ren}}, \ and\ \bibinfo {author} {\bibfnamefont {P.~G.}\ \bibnamefont
  {Kwiat}},\ }\href {\doibase 10.1364/oe.24.010733} {\bibfield  {journal}
  {\bibinfo  {journal} {Optics Express}\ }\textbf {\bibinfo {volume} {24}},\
  \bibinfo {pages} {10733} (\bibinfo {year} {2016})}\BibitemShut {NoStop}%
\bibitem [{\citenamefont {Thiering}\ and\ \citenamefont
  {Gali}(2018{\natexlab{a}})}]{thiering_ab_2018}%
  \BibitemOpen
  \bibfield  {author} {\bibinfo {author} {\bibfnamefont {G.}~\bibnamefont
  {Thiering}}\ and\ \bibinfo {author} {\bibfnamefont {A.}~\bibnamefont
  {Gali}},\ }\href {\doibase 10.1103/PhysRevX.8.021063} {\bibfield  {journal}
  {\bibinfo  {journal} {Phys. Rev. X}\ }\textbf {\bibinfo {volume} {8}},\
  \bibinfo {pages} {021063} (\bibinfo {year} {2018}{\natexlab{a}})}\BibitemShut
  {NoStop}%
\bibitem [{\citenamefont {Knaut}\ \emph {et~al.}(2024)\citenamefont {Knaut},
  \citenamefont {Suleymanzade}, \citenamefont {Wei}, \citenamefont {Assumpcao},
  \citenamefont {Stas}, \citenamefont {Huan}, \citenamefont {Machielse} \emph
  {et~al.}}]{Knaut2024}%
  \BibitemOpen
  \bibfield  {author} {\bibinfo {author} {\bibfnamefont {C.~M.}\ \bibnamefont
  {Knaut}}, \bibinfo {author} {\bibfnamefont {A.}~\bibnamefont {Suleymanzade}},
  \bibinfo {author} {\bibfnamefont {Y.-C.}\ \bibnamefont {Wei}}, \bibinfo
  {author} {\bibfnamefont {D.~R.}\ \bibnamefont {Assumpcao}}, \bibinfo {author}
  {\bibfnamefont {P.-J.}\ \bibnamefont {Stas}}, \bibinfo {author}
  {\bibfnamefont {Y.~Q.}\ \bibnamefont {Huan}}, \bibinfo {author}
  {\bibfnamefont {B.}~\bibnamefont {Machielse}},  \emph {et~al.},\ }\href
  {\doibase 10.1038/s41586-024-07252-z} {\bibfield  {journal} {\bibinfo
  {journal} {Nature}\ }\textbf {\bibinfo {volume} {629}},\ \bibinfo {pages}
  {573} (\bibinfo {year} {2024})}\BibitemShut {NoStop}%
\bibitem [{\citenamefont {Chen}\ \emph {et~al.}(2024)\citenamefont {Chen},
  \citenamefont {Christen}, \citenamefont {Raniwala}, \citenamefont
  {Colangelo}, \citenamefont {Santis}, \citenamefont {Shtyrkova}, \citenamefont
  {Starling}, \citenamefont {Murphy}, \citenamefont {Li}, \citenamefont
  {Berggren}, \citenamefont {Dixon}, \citenamefont {Trusheim},\ and\
  \citenamefont {Englund}}]{chen_scalable_2024}%
  \BibitemOpen
  \bibfield  {author} {\bibinfo {author} {\bibfnamefont {K.~C.}\ \bibnamefont
  {Chen}}, \bibinfo {author} {\bibfnamefont {I.}~\bibnamefont {Christen}},
  \bibinfo {author} {\bibfnamefont {H.}~\bibnamefont {Raniwala}}, \bibinfo
  {author} {\bibfnamefont {M.}~\bibnamefont {Colangelo}}, \bibinfo {author}
  {\bibfnamefont {L.~D.}\ \bibnamefont {Santis}}, \bibinfo {author}
  {\bibfnamefont {K.}~\bibnamefont {Shtyrkova}}, \bibinfo {author}
  {\bibfnamefont {D.}~\bibnamefont {Starling}}, \bibinfo {author}
  {\bibfnamefont {R.}~\bibnamefont {Murphy}}, \bibinfo {author} {\bibfnamefont
  {L.}~\bibnamefont {Li}}, \bibinfo {author} {\bibfnamefont {K.}~\bibnamefont
  {Berggren}}, \bibinfo {author} {\bibfnamefont {P.~B.}\ \bibnamefont {Dixon}},
  \bibinfo {author} {\bibfnamefont {M.}~\bibnamefont {Trusheim}}, \ and\
  \bibinfo {author} {\bibfnamefont {D.}~\bibnamefont {Englund}},\ }\href
  {\doibase 10.1364/OPTICAQ.509233} {\bibfield  {journal} {\bibinfo  {journal}
  {Opt. Quantum}\ }\textbf {\bibinfo {volume} {2}},\ \bibinfo {pages} {124}
  (\bibinfo {year} {2024})}\BibitemShut {NoStop}%
\bibitem [{\citenamefont {Parker}\ \emph {et~al.}(2024)\citenamefont {Parker},
  \citenamefont {Arjona~Martínez}, \citenamefont {Chen}, \citenamefont
  {Stramma}, \citenamefont {Harris}, \citenamefont {Michaels}, \citenamefont
  {Trusheim}, \citenamefont {Hayhurst~Appel}, \citenamefont {Purser},
  \citenamefont {Roth}, \citenamefont {Englund},\ and\ \citenamefont
  {Atatüre}}]{parker_diamond_2024}%
  \BibitemOpen
  \bibfield  {author} {\bibinfo {author} {\bibfnamefont {R.~A.}\ \bibnamefont
  {Parker}}, \bibinfo {author} {\bibfnamefont {J.}~\bibnamefont
  {Arjona~Martínez}}, \bibinfo {author} {\bibfnamefont {K.~C.}\ \bibnamefont
  {Chen}}, \bibinfo {author} {\bibfnamefont {A.~M.}\ \bibnamefont {Stramma}},
  \bibinfo {author} {\bibfnamefont {I.~B.}\ \bibnamefont {Harris}}, \bibinfo
  {author} {\bibfnamefont {C.~P.}\ \bibnamefont {Michaels}}, \bibinfo {author}
  {\bibfnamefont {M.~E.}\ \bibnamefont {Trusheim}}, \bibinfo {author}
  {\bibfnamefont {M.}~\bibnamefont {Hayhurst~Appel}}, \bibinfo {author}
  {\bibfnamefont {C.~M.}\ \bibnamefont {Purser}}, \bibinfo {author}
  {\bibfnamefont {W.~G.}\ \bibnamefont {Roth}}, \bibinfo {author}
  {\bibfnamefont {D.}~\bibnamefont {Englund}}, \ and\ \bibinfo {author}
  {\bibfnamefont {M.}~\bibnamefont {Atatüre}},\ }\href
  {https://www.nature.com/articles/s41566-023-01332-8} {\bibfield  {journal}
  {\bibinfo  {journal} {Nat. Photonics}\ }\textbf {\bibinfo {volume} {18}},\
  \bibinfo {pages} {156} (\bibinfo {year} {2024})}\BibitemShut {NoStop}%
\bibitem [{\citenamefont {Bradac}\ \emph {et~al.}(2019)\citenamefont {Bradac},
  \citenamefont {Gao}, \citenamefont {Forneris}, \citenamefont {Trusheim},\
  and\ \citenamefont {Aharonovich}}]{Bradac2019}%
  \BibitemOpen
  \bibfield  {author} {\bibinfo {author} {\bibfnamefont {C.}~\bibnamefont
  {Bradac}}, \bibinfo {author} {\bibfnamefont {W.}~\bibnamefont {Gao}},
  \bibinfo {author} {\bibfnamefont {J.}~\bibnamefont {Forneris}}, \bibinfo
  {author} {\bibfnamefont {M.~E.}\ \bibnamefont {Trusheim}}, \ and\ \bibinfo
  {author} {\bibfnamefont {I.}~\bibnamefont {Aharonovich}},\ }\href
  {http://dx.doi.org/10.1038/s41467-019-13332-w} {\bibfield  {journal}
  {\bibinfo  {journal} {Nat. Commun.}\ }\textbf {\bibinfo {volume} {10}}
  (\bibinfo {year} {2019})}\BibitemShut {NoStop}%
\bibitem [{\citenamefont {Pieplow}, \citenamefont {Belhassen},\ and\
  \citenamefont {Schröder}(2024)}]{pieplow_efficient_2024}%
  \BibitemOpen
  \bibfield  {author} {\bibinfo {author} {\bibfnamefont {G.}~\bibnamefont
  {Pieplow}}, \bibinfo {author} {\bibfnamefont {M.}~\bibnamefont {Belhassen}},
  \ and\ \bibinfo {author} {\bibfnamefont {T.}~\bibnamefont {Schröder}},\
  }\href {\doibase 10.1103/PhysRevB.109.115409} {\bibfield  {journal} {\bibinfo
   {journal} {Phys. Rev. B}\ }\textbf {\bibinfo {volume} {109}},\ \bibinfo
  {pages} {115409} (\bibinfo {year} {2024})}\BibitemShut {NoStop}%
\bibitem [{\citenamefont {Strocka}\ \emph
  {et~al.}(2025{\natexlab{b}})\citenamefont {Strocka}, \citenamefont
  {Belhassen}, \citenamefont {Schröder},\ and\ \citenamefont
  {Pieplow}}]{code_strocka_qm}%
  \BibitemOpen
  \bibfield  {author} {\bibinfo {author} {\bibfnamefont {Y.}~\bibnamefont
  {Strocka}}, \bibinfo {author} {\bibfnamefont {M.}~\bibnamefont {Belhassen}},
  \bibinfo {author} {\bibfnamefont {T.}~\bibnamefont {Schröder}}, \ and\
  \bibinfo {author} {\bibfnamefont {G.}~\bibnamefont {Pieplow}},\ }\href@noop
  {} {\enquote {\bibinfo {title} {Hybrid quantum memory based on a
  single-photon source and a group-iv color center in diamond},}\ }\bibinfo
  {howpublished}
  {\url{https://github.com/Integrated-Quantum-Photonics-Group/special_issue_adv_qt_v02}}
  (\bibinfo {year} {2025}{\natexlab{b}}),\ \bibinfo {note} {accessed:
  2025-08-14}\BibitemShut {NoStop}%
\bibitem [{\citenamefont {Pieplow}\ \emph {et~al.}(2023)\citenamefont
  {Pieplow}, \citenamefont {Strocka}, \citenamefont {Isaza-Monsalve},
  \citenamefont {Munns},\ and\ \citenamefont
  {Schröder}}]{pieplow_deterministic_2023}%
  \BibitemOpen
  \bibfield  {author} {\bibinfo {author} {\bibfnamefont {G.}~\bibnamefont
  {Pieplow}}, \bibinfo {author} {\bibfnamefont {Y.}~\bibnamefont {Strocka}},
  \bibinfo {author} {\bibfnamefont {M.}~\bibnamefont {Isaza-Monsalve}},
  \bibinfo {author} {\bibfnamefont {J.~H.~D.}\ \bibnamefont {Munns}}, \ and\
  \bibinfo {author} {\bibfnamefont {T.}~\bibnamefont {Schröder}},\ }\href
  {http://arxiv.org/abs/2312.03952} {\bibfield  {journal} {\bibinfo  {journal}
  {arXiv:2312.03952}\ } (\bibinfo {year} {2023})}\BibitemShut {NoStop}%
\bibitem [{\citenamefont {Johnston}(2024)}]{johnston}%
  \BibitemOpen
  \bibfield  {author} {\bibinfo {author} {\bibfnamefont {N.}~\bibnamefont
  {Johnston}},\ }\href
  {https://njohnston.ca/2009/10/the-equivalences-of-the-choi-jamiolkowski-isomorphism-part-i/.}
  {\enquote {\bibinfo {title} {The equivalences of the choi-jamiolkowski
  isomorphism (part i)},}\ } (\bibinfo {year} {2024})\BibitemShut {NoStop}%
\bibitem [{\citenamefont {Sukachev}\ \emph {et~al.}(2017)\citenamefont
  {Sukachev}, \citenamefont {Sipahigil}, \citenamefont {Nguyen}, \citenamefont
  {Bhaskar}, \citenamefont {Evans}, \citenamefont {Jelezko},\ and\
  \citenamefont {Lukin}}]{Sukachev2017}%
  \BibitemOpen
  \bibfield  {author} {\bibinfo {author} {\bibfnamefont {D.}~\bibnamefont
  {Sukachev}}, \bibinfo {author} {\bibfnamefont {A.}~\bibnamefont {Sipahigil}},
  \bibinfo {author} {\bibfnamefont {C.}~\bibnamefont {Nguyen}}, \bibinfo
  {author} {\bibfnamefont {M.}~\bibnamefont {Bhaskar}}, \bibinfo {author}
  {\bibfnamefont {R.}~\bibnamefont {Evans}}, \bibinfo {author} {\bibfnamefont
  {F.}~\bibnamefont {Jelezko}}, \ and\ \bibinfo {author} {\bibfnamefont
  {M.}~\bibnamefont {Lukin}},\ }\href
  {http://dx.doi.org/10.1103/PhysRevLett.119.223602} {\bibfield  {journal}
  {\bibinfo  {journal} {Phys. Rev. Lett.}\ }\textbf {\bibinfo {volume} {119}}
  (\bibinfo {year} {2017})}\BibitemShut {NoStop}%
\bibitem [{\citenamefont {Becker}\ \emph {et~al.}(2016)\citenamefont {Becker},
  \citenamefont {G\"{o}rlitz}, \citenamefont {Arend}, \citenamefont {Markham},\
  and\ \citenamefont {Becher}}]{Becker2016}%
  \BibitemOpen
  \bibfield  {author} {\bibinfo {author} {\bibfnamefont {J.~N.}\ \bibnamefont
  {Becker}}, \bibinfo {author} {\bibfnamefont {J.}~\bibnamefont {G\"{o}rlitz}},
  \bibinfo {author} {\bibfnamefont {C.}~\bibnamefont {Arend}}, \bibinfo
  {author} {\bibfnamefont {M.}~\bibnamefont {Markham}}, \ and\ \bibinfo
  {author} {\bibfnamefont {C.}~\bibnamefont {Becher}},\ }\href
  {https://doi.org/10.1038/ncomms13512} {\bibfield  {journal} {\bibinfo
  {journal} {Nat. Commun.}\ }\textbf {\bibinfo {volume} {7}} (\bibinfo {year}
  {2016})}\BibitemShut {NoStop}%
\bibitem [{\citenamefont {Rosenthal}\ \emph {et~al.}(2023)\citenamefont
  {Rosenthal}, \citenamefont {Anderson}, \citenamefont {Kleidermacher},
  \citenamefont {Stein}, \citenamefont {Lee}, \citenamefont {Grzesik},
  \citenamefont {Scuri}, \citenamefont {Rugar}, \citenamefont {Riedel},
  \citenamefont {Aghaeimeibodi}, \citenamefont {Ahn}, \citenamefont
  {Van~Gasse},\ and\ \citenamefont {Vučković}}]{Rosenthal2023}%
  \BibitemOpen
  \bibfield  {author} {\bibinfo {author} {\bibfnamefont {E.~I.}\ \bibnamefont
  {Rosenthal}}, \bibinfo {author} {\bibfnamefont {C.~P.}\ \bibnamefont
  {Anderson}}, \bibinfo {author} {\bibfnamefont {H.~C.}\ \bibnamefont
  {Kleidermacher}}, \bibinfo {author} {\bibfnamefont {A.~J.}\ \bibnamefont
  {Stein}}, \bibinfo {author} {\bibfnamefont {H.}~\bibnamefont {Lee}}, \bibinfo
  {author} {\bibfnamefont {J.}~\bibnamefont {Grzesik}}, \bibinfo {author}
  {\bibfnamefont {G.}~\bibnamefont {Scuri}}, \bibinfo {author} {\bibfnamefont
  {A.~E.}\ \bibnamefont {Rugar}}, \bibinfo {author} {\bibfnamefont
  {D.}~\bibnamefont {Riedel}}, \bibinfo {author} {\bibfnamefont
  {S.}~\bibnamefont {Aghaeimeibodi}}, \bibinfo {author} {\bibfnamefont {G.~H.}\
  \bibnamefont {Ahn}}, \bibinfo {author} {\bibfnamefont {K.}~\bibnamefont
  {Van~Gasse}}, \ and\ \bibinfo {author} {\bibfnamefont {J.}~\bibnamefont
  {Vučković}},\ }\href {http://dx.doi.org/10.1103/PhysRevX.13.031022}
  {\bibfield  {journal} {\bibinfo  {journal} {Phys. Rev. X}\ }\textbf {\bibinfo
  {volume} {13}} (\bibinfo {year} {2023})}\BibitemShut {NoStop}%
\bibitem [{\citenamefont {Debroux}\ \emph {et~al.}(2021)\citenamefont
  {Debroux}, \citenamefont {Michaels}, \citenamefont {Purser}, \citenamefont
  {Wan}, \citenamefont {Trusheim}, \citenamefont {Mart{\'{\i}}nez},
  \citenamefont {Parker}, \citenamefont {Stramma}, \citenamefont {Chen},
  \citenamefont {de~Santis}, \citenamefont {Alexeev}, \citenamefont {Ferrari},
  \citenamefont {Englund}, \citenamefont {Gangloff},\ and\ \citenamefont
  {Atat\"{u}re}}]{Debroux2021}%
  \BibitemOpen
  \bibfield  {author} {\bibinfo {author} {\bibfnamefont {R.}~\bibnamefont
  {Debroux}}, \bibinfo {author} {\bibfnamefont {C.~P.}\ \bibnamefont
  {Michaels}}, \bibinfo {author} {\bibfnamefont {C.~M.}\ \bibnamefont
  {Purser}}, \bibinfo {author} {\bibfnamefont {N.}~\bibnamefont {Wan}},
  \bibinfo {author} {\bibfnamefont {M.~E.}\ \bibnamefont {Trusheim}}, \bibinfo
  {author} {\bibfnamefont {J.~A.}\ \bibnamefont {Mart{\'{\i}}nez}}, \bibinfo
  {author} {\bibfnamefont {R.~A.}\ \bibnamefont {Parker}}, \bibinfo {author}
  {\bibfnamefont {A.~M.}\ \bibnamefont {Stramma}}, \bibinfo {author}
  {\bibfnamefont {K.~C.}\ \bibnamefont {Chen}}, \bibinfo {author}
  {\bibfnamefont {L.}~\bibnamefont {de~Santis}}, \bibinfo {author}
  {\bibfnamefont {E.~M.}\ \bibnamefont {Alexeev}}, \bibinfo {author}
  {\bibfnamefont {A.~C.}\ \bibnamefont {Ferrari}}, \bibinfo {author}
  {\bibfnamefont {D.}~\bibnamefont {Englund}}, \bibinfo {author} {\bibfnamefont
  {D.~A.}\ \bibnamefont {Gangloff}}, \ and\ \bibinfo {author} {\bibfnamefont
  {M.}~\bibnamefont {Atat\"{u}re}},\ }\href
  {https://doi.org/10.1103/physrevx.11.041041} {\bibfield  {journal} {\bibinfo
  {journal} {Phys. Rev. X}\ }\textbf {\bibinfo {volume} {11}} (\bibinfo {year}
  {2021})}\BibitemShut {NoStop}%
\bibitem [{\citenamefont {Omlor}, \citenamefont {Tissot},\ and\ \citenamefont
  {Burkard}(2025)}]{omlor_entanglement_2024}%
  \BibitemOpen
  \bibfield  {author} {\bibinfo {author} {\bibfnamefont {F.}~\bibnamefont
  {Omlor}}, \bibinfo {author} {\bibfnamefont {B.}~\bibnamefont {Tissot}}, \
  and\ \bibinfo {author} {\bibfnamefont {G.}~\bibnamefont {Burkard}},\ }\href
  {http://dx.doi.org/10.1103/PhysRevA.111.012612} {\bibfield  {journal}
  {\bibinfo  {journal} {Phys. Rev. A}\ }\textbf {\bibinfo {volume} {111}}
  (\bibinfo {year} {2025})}\BibitemShut {NoStop}%
\bibitem [{\citenamefont {Podlecki}, \citenamefont {Martin},\ and\
  \citenamefont {Bastin}(2021)}]{Podlecki2021}%
  \BibitemOpen
  \bibfield  {author} {\bibinfo {author} {\bibfnamefont {L.}~\bibnamefont
  {Podlecki}}, \bibinfo {author} {\bibfnamefont {J.}~\bibnamefont {Martin}}, \
  and\ \bibinfo {author} {\bibfnamefont {T.}~\bibnamefont {Bastin}},\ }\href
  {http://dx.doi.org/10.1364/JOSAB.433090} {\bibfield  {journal} {\bibinfo
  {journal} {J. Opt. Soc. Am. B}\ }\textbf {\bibinfo {volume} {38}},\ \bibinfo
  {pages} {3244} (\bibinfo {year} {2021})}\BibitemShut {NoStop}%
\bibitem [{\citenamefont {Reiserer}\ and\ \citenamefont
  {Rempe}(2015)}]{reiserer_cavity-based_2015}%
  \BibitemOpen
  \bibfield  {author} {\bibinfo {author} {\bibfnamefont {A.}~\bibnamefont
  {Reiserer}}\ and\ \bibinfo {author} {\bibfnamefont {G.}~\bibnamefont
  {Rempe}},\ }\href {\doibase 10.1103/RevModPhys.87.1379} {\bibfield  {journal}
  {\bibinfo  {journal} {Rev. Mod. Phys.}\ }\textbf {\bibinfo {volume} {87}},\
  \bibinfo {pages} {1379} (\bibinfo {year} {2015})}\BibitemShut {NoStop}%
\bibitem [{\citenamefont {Bhaskar}\ \emph
  {et~al.}(2020{\natexlab{b}})\citenamefont {Bhaskar}, \citenamefont
  {Riedinger}, \citenamefont {Machielse}, \citenamefont {Levonian},
  \citenamefont {Nguyen}, \citenamefont {Knall}, \citenamefont {Park},
  \citenamefont {Englund}, \citenamefont {Lončar}, \citenamefont {Sukachev},\
  and\ \citenamefont {Lukin}}]{bhaskar_experimental_2020}%
  \BibitemOpen
  \bibfield  {author} {\bibinfo {author} {\bibfnamefont {M.~K.}\ \bibnamefont
  {Bhaskar}}, \bibinfo {author} {\bibfnamefont {R.}~\bibnamefont {Riedinger}},
  \bibinfo {author} {\bibfnamefont {B.}~\bibnamefont {Machielse}}, \bibinfo
  {author} {\bibfnamefont {D.~S.}\ \bibnamefont {Levonian}}, \bibinfo {author}
  {\bibfnamefont {C.~T.}\ \bibnamefont {Nguyen}}, \bibinfo {author}
  {\bibfnamefont {E.~N.}\ \bibnamefont {Knall}}, \bibinfo {author}
  {\bibfnamefont {H.}~\bibnamefont {Park}}, \bibinfo {author} {\bibfnamefont
  {D.}~\bibnamefont {Englund}}, \bibinfo {author} {\bibfnamefont
  {M.}~\bibnamefont {Lončar}}, \bibinfo {author} {\bibfnamefont {D.~D.}\
  \bibnamefont {Sukachev}}, \ and\ \bibinfo {author} {\bibfnamefont {M.~D.}\
  \bibnamefont {Lukin}},\ }\href {\doibase 10.1038/s41586-020-2103-5}
  {\bibfield  {journal} {\bibinfo  {journal} {Nature}\ }\textbf {\bibinfo
  {volume} {580}},\ \bibinfo {pages} {60} (\bibinfo {year}
  {2020}{\natexlab{b}})}\BibitemShut {NoStop}%
\bibitem [{\citenamefont {Herrmann}\ \emph {et~al.}(2024)\citenamefont
  {Herrmann}, \citenamefont {Fischer}, \citenamefont {Brevoord}, \citenamefont
  {Sauerzapf}, \citenamefont {Wienhoven}, \citenamefont {Feije}, \citenamefont
  {Pasini}, \citenamefont {Eschen}, \citenamefont {Ruf}, \citenamefont
  {Weaver},\ and\ \citenamefont {Hanson}}]{Herrmann2024}%
  \BibitemOpen
  \bibfield  {author} {\bibinfo {author} {\bibfnamefont {Y.}~\bibnamefont
  {Herrmann}}, \bibinfo {author} {\bibfnamefont {J.}~\bibnamefont {Fischer}},
  \bibinfo {author} {\bibfnamefont {J.~M.}\ \bibnamefont {Brevoord}}, \bibinfo
  {author} {\bibfnamefont {C.}~\bibnamefont {Sauerzapf}}, \bibinfo {author}
  {\bibfnamefont {L.~G.}\ \bibnamefont {Wienhoven}}, \bibinfo {author}
  {\bibfnamefont {L.~J.}\ \bibnamefont {Feije}}, \bibinfo {author}
  {\bibfnamefont {M.}~\bibnamefont {Pasini}}, \bibinfo {author} {\bibfnamefont
  {M.}~\bibnamefont {Eschen}}, \bibinfo {author} {\bibfnamefont
  {M.}~\bibnamefont {Ruf}}, \bibinfo {author} {\bibfnamefont {M.~J.}\
  \bibnamefont {Weaver}}, \ and\ \bibinfo {author} {\bibfnamefont
  {R.}~\bibnamefont {Hanson}},\ }\href
  {http://dx.doi.org/10.1103/PhysRevX.14.041013} {\bibfield  {journal}
  {\bibinfo  {journal} {Phys. Rev. X}\ }\textbf {\bibinfo {volume} {14}}
  (\bibinfo {year} {2024})}\BibitemShut {NoStop}%
\bibitem [{\citenamefont {Cherednichenko}\ \emph {et~al.}(2021)\citenamefont
  {Cherednichenko}, \citenamefont {Acharya}, \citenamefont {Novoselov},\ and\
  \citenamefont {Drakinskiy}}]{cherednichenko_low_2021}%
  \BibitemOpen
  \bibfield  {author} {\bibinfo {author} {\bibfnamefont {S.}~\bibnamefont
  {Cherednichenko}}, \bibinfo {author} {\bibfnamefont {N.}~\bibnamefont
  {Acharya}}, \bibinfo {author} {\bibfnamefont {E.}~\bibnamefont {Novoselov}},
  \ and\ \bibinfo {author} {\bibfnamefont {V.}~\bibnamefont {Drakinskiy}},\
  }\href {http://arxiv.org/abs/1911.01480} {\bibfield  {journal} {\bibinfo
  {journal} {Supercond. Sci. Technol.}\ }\textbf {\bibinfo {volume} {34}},\
  \bibinfo {pages} {044001} (\bibinfo {year} {2021})}\BibitemShut {NoStop}%
\bibitem [{\citenamefont {Grotowski}\ \emph {et~al.}(2025)\citenamefont
  {Grotowski}, \citenamefont {Zugliani}, \citenamefont {Jonas}, \citenamefont
  {Flaschmann}, \citenamefont {Schmid}, \citenamefont {Strohauer},
  \citenamefont {Wietschorke}, \citenamefont {Bruckmoser}, \citenamefont
  {M\"{u}ller}, \citenamefont {Althammer}, \citenamefont {Gross}, \citenamefont
  {M\"{u}ller},\ and\ \citenamefont {Finley}}]{Grotowski2025}%
  \BibitemOpen
  \bibfield  {author} {\bibinfo {author} {\bibfnamefont {S.}~\bibnamefont
  {Grotowski}}, \bibinfo {author} {\bibfnamefont {L.}~\bibnamefont {Zugliani}},
  \bibinfo {author} {\bibfnamefont {B.}~\bibnamefont {Jonas}}, \bibinfo
  {author} {\bibfnamefont {R.}~\bibnamefont {Flaschmann}}, \bibinfo {author}
  {\bibfnamefont {C.}~\bibnamefont {Schmid}}, \bibinfo {author} {\bibfnamefont
  {S.}~\bibnamefont {Strohauer}}, \bibinfo {author} {\bibfnamefont
  {F.}~\bibnamefont {Wietschorke}}, \bibinfo {author} {\bibfnamefont
  {N.}~\bibnamefont {Bruckmoser}}, \bibinfo {author} {\bibfnamefont
  {M.}~\bibnamefont {M\"{u}ller}}, \bibinfo {author} {\bibfnamefont
  {M.}~\bibnamefont {Althammer}}, \bibinfo {author} {\bibfnamefont
  {R.}~\bibnamefont {Gross}}, \bibinfo {author} {\bibfnamefont
  {K.}~\bibnamefont {M\"{u}ller}}, \ and\ \bibinfo {author} {\bibfnamefont
  {J.}~\bibnamefont {Finley}},\ }\href
  {http://dx.doi.org/10.1038/s41598-025-86303-5} {\bibfield  {journal}
  {\bibinfo  {journal} {Sci. Rep.}\ }\textbf {\bibinfo {volume} {15}} (\bibinfo
  {year} {2025})}\BibitemShut {NoStop}%
\bibitem [{\citenamefont {Tran}\ \emph {et~al.}(2017)\citenamefont {Tran},
  \citenamefont {Kianinia}, \citenamefont {Bray}, \citenamefont {Kim},
  \citenamefont {Xu}, \citenamefont {Gentle}, \citenamefont {Sontheimer},
  \citenamefont {Bradac},\ and\ \citenamefont
  {Aharonovich}}]{tran_nanodiamonds_2017}%
  \BibitemOpen
  \bibfield  {author} {\bibinfo {author} {\bibfnamefont {T.~T.}\ \bibnamefont
  {Tran}}, \bibinfo {author} {\bibfnamefont {M.}~\bibnamefont {Kianinia}},
  \bibinfo {author} {\bibfnamefont {K.}~\bibnamefont {Bray}}, \bibinfo {author}
  {\bibfnamefont {S.}~\bibnamefont {Kim}}, \bibinfo {author} {\bibfnamefont
  {Z.-Q.}\ \bibnamefont {Xu}}, \bibinfo {author} {\bibfnamefont
  {A.}~\bibnamefont {Gentle}}, \bibinfo {author} {\bibfnamefont
  {B.}~\bibnamefont {Sontheimer}}, \bibinfo {author} {\bibfnamefont
  {C.}~\bibnamefont {Bradac}}, \ and\ \bibinfo {author} {\bibfnamefont
  {I.}~\bibnamefont {Aharonovich}},\ }\href {\doibase 10.1063/1.4998199}
  {\bibfield  {journal} {\bibinfo  {journal} {APL Photonics}\ }\textbf
  {\bibinfo {volume} {2}},\ \bibinfo {pages} {116103} (\bibinfo {year}
  {2017})}\BibitemShut {NoStop}%
\bibitem [{\citenamefont {Tiurev}\ \emph {et~al.}(2021)\citenamefont {Tiurev},
  \citenamefont {Mirambell}, \citenamefont {Lauritzen}, \citenamefont {Appel},
  \citenamefont {Tiranov}, \citenamefont {Lodahl},\ and\ \citenamefont
  {S{\o}rensen}}]{tiurev2021fidelity}%
  \BibitemOpen
  \bibfield  {author} {\bibinfo {author} {\bibfnamefont {K.}~\bibnamefont
  {Tiurev}}, \bibinfo {author} {\bibfnamefont {P.~L.}\ \bibnamefont
  {Mirambell}}, \bibinfo {author} {\bibfnamefont {M.~B.}\ \bibnamefont
  {Lauritzen}}, \bibinfo {author} {\bibfnamefont {M.~H.}\ \bibnamefont
  {Appel}}, \bibinfo {author} {\bibfnamefont {A.}~\bibnamefont {Tiranov}},
  \bibinfo {author} {\bibfnamefont {P.}~\bibnamefont {Lodahl}}, \ and\ \bibinfo
  {author} {\bibfnamefont {A.~S.}\ \bibnamefont {S{\o}rensen}},\ }\href
  {\doibase 10.1103/PhysRevA.104.052604} {\bibfield  {journal} {\bibinfo
  {journal} {Phys. Rev. A}\ }\textbf {\bibinfo {volume} {104}},\ \bibinfo
  {pages} {052604} (\bibinfo {year} {2021})}\BibitemShut {NoStop}%
\bibitem [{\citenamefont {Liang}\ \emph {et~al.}(2019)\citenamefont {Liang},
  \citenamefont {Yeh}, \citenamefont {Mendon\c{c}a}, \citenamefont {Teh},
  \citenamefont {Reid},\ and\ \citenamefont {Drummond}}]{Liang2019}%
  \BibitemOpen
  \bibfield  {author} {\bibinfo {author} {\bibfnamefont {Y.-C.}\ \bibnamefont
  {Liang}}, \bibinfo {author} {\bibfnamefont {Y.-H.}\ \bibnamefont {Yeh}},
  \bibinfo {author} {\bibfnamefont {P.~E. M.~F.}\ \bibnamefont {Mendon\c{c}a}},
  \bibinfo {author} {\bibfnamefont {R.~Y.}\ \bibnamefont {Teh}}, \bibinfo
  {author} {\bibfnamefont {M.~D.}\ \bibnamefont {Reid}}, \ and\ \bibinfo
  {author} {\bibfnamefont {P.~D.}\ \bibnamefont {Drummond}},\ }\href
  {http://dx.doi.org/10.1088/1361-6633/ab1ca4} {\bibfield  {journal} {\bibinfo
  {journal} {Rep. Prog. Phys.}\ }\textbf {\bibinfo {volume} {82}},\ \bibinfo
  {pages} {076001} (\bibinfo {year} {2019})}\BibitemShut {NoStop}%
\bibitem [{\citenamefont {Endres}, \citenamefont {Sandrock},\ and\
  \citenamefont {Focke}(2018)}]{endres_simplicial_2018}%
  \BibitemOpen
  \bibfield  {author} {\bibinfo {author} {\bibfnamefont {S.~C.}\ \bibnamefont
  {Endres}}, \bibinfo {author} {\bibfnamefont {C.}~\bibnamefont {Sandrock}}, \
  and\ \bibinfo {author} {\bibfnamefont {W.~W.}\ \bibnamefont {Focke}},\ }\href
  {\doibase 10.1007/s10898-018-0645-y} {\bibfield  {journal} {\bibinfo
  {journal} {J. Glob. Optim.}\ }\textbf {\bibinfo {volume} {72}},\ \bibinfo
  {pages} {181} (\bibinfo {year} {2018})}\BibitemShut {NoStop}%
\bibitem [{\citenamefont {et~al.}(2020)}]{2020SciPy-NMeth}%
  \BibitemOpen
  \bibfield  {author} {\bibinfo {author} {\bibfnamefont {P.~V.}\ \bibnamefont
  {et~al.}},\ }\href {\doibase 10.1038/s41592-019-0686-2} {\bibfield  {journal}
  {\bibinfo  {journal} {Nat. Methods}\ }\textbf {\bibinfo {volume} {17}},\
  \bibinfo {pages} {261} (\bibinfo {year} {2020})}\BibitemShut {NoStop}%
\bibitem [{\citenamefont {Wang}(2021)}]{Wang2021}%
  \BibitemOpen
  \bibfield  {author} {\bibinfo {author} {\bibfnamefont {D.}~\bibnamefont
  {Wang}},\ }\href {http://dx.doi.org/10.1088/1361-6455/abf6e1} {\bibfield
  {journal} {\bibinfo  {journal} {J. Phys. B: At. Mol. Opt. Phys.}\ }\textbf
  {\bibinfo {volume} {54}},\ \bibinfo {pages} {133001} (\bibinfo {year}
  {2021})}\BibitemShut {NoStop}%
\bibitem [{\citenamefont {Bongioanni}\ \emph {et~al.}(2010)\citenamefont
  {Bongioanni}, \citenamefont {Sansoni}, \citenamefont {Sciarrino},
  \citenamefont {Vallone},\ and\ \citenamefont {Mataloni}}]{Bongioanni2010}%
  \BibitemOpen
  \bibfield  {author} {\bibinfo {author} {\bibfnamefont {I.}~\bibnamefont
  {Bongioanni}}, \bibinfo {author} {\bibfnamefont {L.}~\bibnamefont {Sansoni}},
  \bibinfo {author} {\bibfnamefont {F.}~\bibnamefont {Sciarrino}}, \bibinfo
  {author} {\bibfnamefont {G.}~\bibnamefont {Vallone}}, \ and\ \bibinfo
  {author} {\bibfnamefont {P.}~\bibnamefont {Mataloni}},\ }\href
  {http://dx.doi.org/10.1103/PhysRevA.82.042307} {\bibfield  {journal}
  {\bibinfo  {journal} {Phys. Rev. A}\ }\textbf {\bibinfo {volume} {82}}
  (\bibinfo {year} {2010})}\BibitemShut {NoStop}%
\bibitem [{\citenamefont {Bhandari}\ and\ \citenamefont
  {Peters}(2016)}]{Bhandari2016}%
  \BibitemOpen
  \bibfield  {author} {\bibinfo {author} {\bibfnamefont {R.}~\bibnamefont
  {Bhandari}}\ and\ \bibinfo {author} {\bibfnamefont {N.~A.}\ \bibnamefont
  {Peters}},\ }\href {http://dx.doi.org/10.1038/srep26004} {\bibfield
  {journal} {\bibinfo  {journal} {Sci. Rep.}\ }\textbf {\bibinfo {volume} {6}}
  (\bibinfo {year} {2016})}\BibitemShut {NoStop}%
\bibitem [{\citenamefont {Dietrich}\ \emph {et~al.}(2020)\citenamefont
  {Dietrich}, \citenamefont {Doherty}, \citenamefont {Aharonovich},\ and\
  \citenamefont {Kubanek}}]{Dietrich2020}%
  \BibitemOpen
  \bibfield  {author} {\bibinfo {author} {\bibfnamefont {A.}~\bibnamefont
  {Dietrich}}, \bibinfo {author} {\bibfnamefont {M.~W.}\ \bibnamefont
  {Doherty}}, \bibinfo {author} {\bibfnamefont {I.}~\bibnamefont
  {Aharonovich}}, \ and\ \bibinfo {author} {\bibfnamefont {A.}~\bibnamefont
  {Kubanek}},\ }\href {http://dx.doi.org/10.1103/PhysRevB.101.081401}
  {\bibfield  {journal} {\bibinfo  {journal} {Phys. Rev. B}\ }\textbf {\bibinfo
  {volume} {101}} (\bibinfo {year} {2020})}\BibitemShut {NoStop}%
\bibitem [{\citenamefont {Strocka}\ \emph
  {et~al.}(2025{\natexlab{c}})\citenamefont {Strocka}, \citenamefont {Gu},
  \citenamefont {Pieplow}, \citenamefont {Borregaard},\ and\ \citenamefont
  {Schr\"{o}der}}]{strocka_repeater_2025}%
  \BibitemOpen
  \bibfield  {author} {\bibinfo {author} {\bibfnamefont {Y.}~\bibnamefont
  {Strocka}}, \bibinfo {author} {\bibfnamefont {F.}~\bibnamefont {Gu}},
  \bibinfo {author} {\bibfnamefont {G.}~\bibnamefont {Pieplow}}, \bibinfo
  {author} {\bibfnamefont {J.}~\bibnamefont {Borregaard}}, \ and\ \bibinfo
  {author} {\bibfnamefont {T.}~\bibnamefont {Schr\"{o}der}},\ }\href
  {https://arxiv.org/abs/2511.02472} {\bibfield  {journal} {\bibinfo  {journal}
  {arXiv:2511.02472}\ } (\bibinfo {year} {2025}{\natexlab{c}})}\BibitemShut
  {NoStop}%
\bibitem [{\citenamefont {Thiering}\ and\ \citenamefont
  {Gali}(2018{\natexlab{b}})}]{Thiering2018}%
  \BibitemOpen
  \bibfield  {author} {\bibinfo {author} {\bibfnamefont {G.}~\bibnamefont
  {Thiering}}\ and\ \bibinfo {author} {\bibfnamefont {A.}~\bibnamefont
  {Gali}},\ }\href {http://dx.doi.org/10.1103/PhysRevX.8.021063} {\bibfield
  {journal} {\bibinfo  {journal} {Phys. Rev. X}\ }\textbf {\bibinfo {volume}
  {8}} (\bibinfo {year} {2018}{\natexlab{b}})}\BibitemShut {NoStop}%
\bibitem [{\citenamefont {Floch}\ \emph {et~al.}(2011)\citenamefont {Floch},
  \citenamefont {Bara}, \citenamefont {Hartnett}, \citenamefont {Tobar},
  \citenamefont {Mouneyrac}, \citenamefont {Passerieux}, \citenamefont {Cros},
  \citenamefont {Krupka}, \citenamefont {Goy},\ and\ \citenamefont
  {Caroopen}}]{floch_electromagnetic_2011}%
  \BibitemOpen
  \bibfield  {author} {\bibinfo {author} {\bibfnamefont {J.-M.~L.}\
  \bibnamefont {Floch}}, \bibinfo {author} {\bibfnamefont {R.}~\bibnamefont
  {Bara}}, \bibinfo {author} {\bibfnamefont {J.~G.}\ \bibnamefont {Hartnett}},
  \bibinfo {author} {\bibfnamefont {M.~E.}\ \bibnamefont {Tobar}}, \bibinfo
  {author} {\bibfnamefont {D.}~\bibnamefont {Mouneyrac}}, \bibinfo {author}
  {\bibfnamefont {D.}~\bibnamefont {Passerieux}}, \bibinfo {author}
  {\bibfnamefont {D.}~\bibnamefont {Cros}}, \bibinfo {author} {\bibfnamefont
  {J.}~\bibnamefont {Krupka}}, \bibinfo {author} {\bibfnamefont
  {P.}~\bibnamefont {Goy}}, \ and\ \bibinfo {author} {\bibfnamefont
  {S.}~\bibnamefont {Caroopen}},\ }\href {\doibase 10.1063/1.3580903}
  {\bibfield  {journal} {\bibinfo  {journal} {J. Appl. Phys.}\ }\textbf
  {\bibinfo {volume} {109}},\ \bibinfo {pages} {094103} (\bibinfo {year}
  {2011})}\BibitemShut {NoStop}%
\bibitem [{\citenamefont {Bayn}\ and\ \citenamefont
  {Salzman}(2008)}]{Bayn2008}%
  \BibitemOpen
  \bibfield  {author} {\bibinfo {author} {\bibfnamefont {I.}~\bibnamefont
  {Bayn}}\ and\ \bibinfo {author} {\bibfnamefont {J.}~\bibnamefont {Salzman}},\
  }\href {\doibase 10.1364/OE.16.004972} {\bibfield  {journal} {\bibinfo
  {journal} {Opt. Express}\ }\textbf {\bibinfo {volume} {16}},\ \bibinfo
  {pages} {4972} (\bibinfo {year} {2008})}\BibitemShut {NoStop}%
\bibitem [{\citenamefont {Cao}\ \emph {et~al.}(2023)\citenamefont {Cao},
  \citenamefont {Yang}, \citenamefont {Fandrich}, \citenamefont {Zhang},
  \citenamefont {Rugeramigabo}, \citenamefont {Brechtken}, \citenamefont
  {Haug}, \citenamefont {Zopf},\ and\ \citenamefont
  {Ding}}]{cao_solid-state_2023}%
  \BibitemOpen
  \bibfield  {author} {\bibinfo {author} {\bibfnamefont {X.}~\bibnamefont
  {Cao}}, \bibinfo {author} {\bibfnamefont {J.}~\bibnamefont {Yang}}, \bibinfo
  {author} {\bibfnamefont {T.}~\bibnamefont {Fandrich}}, \bibinfo {author}
  {\bibfnamefont {Y.}~\bibnamefont {Zhang}}, \bibinfo {author} {\bibfnamefont
  {E.~P.}\ \bibnamefont {Rugeramigabo}}, \bibinfo {author} {\bibfnamefont
  {B.}~\bibnamefont {Brechtken}}, \bibinfo {author} {\bibfnamefont {R.~J.}\
  \bibnamefont {Haug}}, \bibinfo {author} {\bibfnamefont {M.}~\bibnamefont
  {Zopf}}, \ and\ \bibinfo {author} {\bibfnamefont {F.}~\bibnamefont {Ding}},\
  }\href {\doibase 10.1021/acs.nanolett.3c01570} {\bibfield  {journal}
  {\bibinfo  {journal} {Nano Lett.}\ }\textbf {\bibinfo {volume} {23}},\
  \bibinfo {pages} {6109} (\bibinfo {year} {2023})}\BibitemShut {NoStop}%
\bibitem [{\citenamefont {Kimble}(1998)}]{kimble_strong_1998}%
  \BibitemOpen
  \bibfield  {author} {\bibinfo {author} {\bibfnamefont {H.~J.}\ \bibnamefont
  {Kimble}},\ }\href {\doibase 10.1238/Physica.Topical.076a00127} {\bibfield
  {journal} {\bibinfo  {journal} {Phys. Scripta}\ }\textbf {\bibinfo {volume}
  {1998}},\ \bibinfo {pages} {127} (\bibinfo {year} {1998})}\BibitemShut
  {NoStop}%
\bibitem [{\citenamefont {Propp}(2022)}]{propp}%
  \BibitemOpen
  \bibfield  {author} {\bibinfo {author} {\bibfnamefont {T.~B.}\ \bibnamefont
  {Propp}},\ }\href {https://arxiv.org/abs/2210.04089} {\bibfield  {journal}
  {\bibinfo  {journal} {arXiv:2210.04089}\ } (\bibinfo {year}
  {2022})}\BibitemShut {NoStop}%
\bibitem [{\citenamefont {Collins}\ and\ \citenamefont
  {Stephens}(2015)}]{collins_depolarizing-channel_2015}%
  \BibitemOpen
  \bibfield  {author} {\bibinfo {author} {\bibfnamefont {D.}~\bibnamefont
  {Collins}}\ and\ \bibinfo {author} {\bibfnamefont {J.}~\bibnamefont
  {Stephens}},\ }\href {\doibase 10.1103/PhysRevA.92.032324} {\bibfield
  {journal} {\bibinfo  {journal} {Phys. Rev. A}\ }\textbf {\bibinfo {volume}
  {92}},\ \bibinfo {pages} {032324} (\bibinfo {year} {2015})}\BibitemShut
  {NoStop}%
\end{thebibliography}%

\onecolumngrid
\begin{appendix}
\section{SPIN-PHOTON INTERACTION}
\label{app:system}

Group-IV color centers (G4V) in diamond have four states $\{\ket{k}\}$ with $k\in G=\{1,2,3,4\}$ in the ground state manifold and four states $\{\ket{l}\}$ with $l\in E=\{A,B,C,D\}$ in the excited state manifold. Therefore, the Hamiltonian of the G4V coupled to a single cavity mode is given by
\begin{align}\label{eq:hamiltonian}
    H=\sum_{m\in G\cup E} \epsilon_m\ket{m}\bra{m}+\omega_c a_c^\dagger a_c+\left(\sum_{k\in G,l\in E}g_{kl}\sigma_{kl}+g_{kl}^*\sigma_{lk} \right)\left(a+a^\dagger\right)
\end{align}
where $\epsilon_m$ is the bare energy of quantum state $\ket{m}$ with $m \in G\cup E$, $\omega_c$ is the central frequency of the cavity mode, and $\sigma_{kl}=\ket{k}\bra{l}$ is the transition operator. Here,
\begin{align}
    g_{kl}={\rm i}\sqrt{\frac{\omega_c}{2\hbar\epsilon_0\epsilon_r V}}\bra{k}\bm{\epsilon}\cdot\mathbf{d}\ket{l}
\end{align}
is the coupling strength between $\ket{k}\leftrightarrow\ket{l}$ transition and the cavity mode, where $\mathbf{\epsilon}$ is the cavity mode orientation which is made parallel to the defect's symmetry axis, i.e. $\bm{\epsilon}=1/\sqrt{3}(1,1,1)$ in the diamond lattice coordinate system, $\mathbf{d}$ is the G4V dipole moment operator, $\epsilon_r=5.7$ is the relative permittivity of diamond~\cite{floch_electromagnetic_2011},  $V=V_{{\rm eff}}\frac{\lambda^3}{2n^3}$ is the mode volume, where $V_{{\rm eff}}=1.8$~\cite{Bayn2008}, $n=2.417$ is the refractive index of diamond~\cite{bhaskar_experimental_2020} and $\lambda=2\pi c/\omega_{c}$ is the wavelength of the cavity mode, where $c$ is the speed of the light.

Here, the cavity-G4V setup is considered as an open system with Lindblad operators
\begin{gather}
    L_1=\sqrt{\gamma_{1A}}\sigma_{1A}\;, \\
    L_2=\sqrt{\gamma_{1B}}\sigma_{1B}\;, \\
    L_3=\sqrt{\gamma_{2A}}\sigma_{2A}\;, \\
    L_4=\sqrt{\gamma_{2B}}\sigma_{2B}\;, \\
    L_5=\sqrt{2\kappa} a\;,
\end{gather}
where $\kappa$ is the photon loss rate of the single-sided cavity (a half-width at half-maximum of the spectral curve), 
\begin{align}
    \gamma_{\alpha\beta}=\frac{4\alpha\omega_{\alpha\beta}^3 n\vert\bra{\alpha} \mathbf{d}\ket{\beta}\vert^2}{3c^2 e^2},
\end{align}
where $\alpha=1, 2$ and $\beta=A, B$, are the G4V natural decay rates based on Fermi's golden rule~\cite{pieplow_deterministic_2023}. Here, $\alpha=1/137$ is the fine structure constant, $e$ is the elementary charge. The G4V natural decay rates relate to the decay time $T_1$ as $\gamma=1/T_1$. For the SiV$^{-}$(SnV), $T_1=1.7 (4.5)$ ns \cite{cao_solid-state_2023, pieplow_deterministic_2023}, with the Debye-Waller factor $0.8 (0.6)$~\cite{bopp_sawfish_2024, pieplow_deterministic_2023}. Additionally, we define the theoretical cooperativities~\cite{kimble_strong_1998} as a function of the coupling strength $g_{kl}$, full width at half maximum of the spectral curve $\gamma_{kl}$, and full width at half maximum of the spectral curve $2\kappa$ as
\begin{align}
    C_{kl}=\frac{\vert g_{kl}\vert^2}{2\kappa\gamma_{k}}.
\end{align}
In \cite{bhaskar_experimental_2020} the experimental cooperativity is used which is $C_{kl,\rm bh}=4C_{kl}$.

In the rotating frame after the unitary transformation of the unitary operator $U(t)=e^{-{\rm i}H_0 t}$, where $H_0=\sum_{m=2}^8 \epsilon_m\ket{m}\bra{m}+\omega_c a_c^\dagger a_c$, the Hamiltonian can be expressed as
\begin{align}
    \tilde{H}(t)=&\sum_{k\in G,l\in E}\Big(e^{{\rm i}(\epsilon_k-\epsilon_l-\omega_c)t}g_{kl}\sigma_{kl}a +e^{{\rm i}(\epsilon_k-\epsilon_l+\omega_c)t}g_{kl}\sigma_{kl}a^\dagger +e^{-{\rm i}(\epsilon_k-\epsilon_l+\omega_c)t}g_{kl}^*\sigma_{lk}a +e^{-{\rm i}(\epsilon_k-\epsilon_l-\omega_c)t}g_{kl}^*\sigma_{lk}a^\dagger\Big)\;.
\end{align}

Neglecting the terms which fulfill $\epsilon_k-\epsilon_l\pm\omega_c\gg{\rm max}\vert \bm{g}\vert$ (rotating-wave approximation) and defining $\delta=\epsilon_A-\omega_c$ and $\omega_s=\epsilon_2$, the Hamilitonian becomes
\begin{align*}
    H_{\rm RWA}(t)=&e^{-{\rm i}\delta t}g_{1A}\sigma_{1A}a^\dagger+e^{{\rm i}(\omega_s-\delta) t}g_{2A}\sigma_{2A}a^\dagger+e^{{\rm i}(\epsilon_A-\epsilon_B-\delta)t}g_{1B}\sigma_{1B}a^\dagger+e^{{\rm i}(\omega_s-\epsilon_B+\epsilon_A-\delta)t}g_{2B}\sigma_{2B}a^\dagger+{\rm H.c.}
\end{align*}

In the frame after another unitary transformation of the unitary operator
\begin{align}
    U(t)=e^{-{\rm i}(\delta_A\ket{A}\bra{A}+\delta_B\ket{B}\bra{B})t}
\end{align} with $\delta_A=-\delta$ and $\delta_B=\omega_s-\epsilon_B+\epsilon_A-\delta=-\omega_{2B}+\omega_c$, the Hamiltonian becomes
\begin{align}\label{eq:cross_ham}
    H_r(t)&=U^\dagger (t) H_{\rm RWA}(t) U(t)- {\rm i}\dot{U}(t)U(t)^\dagger\\
    &=H'_0+g_{1A}\sigma_{1A}a^\dagger+e^{{\rm i}\omega_s t}g_{2A}\sigma_{2A} a^\dagger +e^{-{\rm i}\omega_s t}g_{1B}\sigma_{1B}a^\dagger+g_{2B}\sigma_{2B}a^\dagger+{\rm H.c.}\;,
\end{align}
where $H'_0=-\delta_A\ket{A}\bra{A}-\delta_B\ket{B}\bra{B}$.
Now consider an input mode $a_{\rm in}$. The driving Hamiltonian is
\begin{align}
    H_{\rm drive}(t)={\rm i}\sqrt{2k_l}(a_{\rm in}(t)a^\dagger-a_{\rm in}^*(t) a)
\end{align}

For an open quantum system with Hamiltonian $H_r$ and Lindblad operators $\{L_i\}$, and in the Heisenberg picture, the dynamics of an operator $X$ is governed by the Heisenberg-Langevin equation
\begin{equation}
    {\dot {X}}={i}[H_r+H_{\rm drive}(t),X]+\sum _{m=1}^5 L^\dagger_{m} XL_{m}-{\frac {1}{2}}\left\{L_{m}^{\dagger }L_{m},X\right\}
\end{equation}
for the operators $X=a,\sigma_{1A},\sigma_{2A},\sigma_{1B},\sigma_{2B},\sigma_{11},\sigma_{22},\sigma_{AA},\sigma_{BB}$.
We end up evaluating the commutators of the equation system
\begin{align}
    \dot{a}&=- {\rm i}[a,H_{{r}}]-\kappa a+\sqrt{2\kappa_l}a_{{\rm in}}(t),\\
    \dot{\sigma}_{1A}&=-{\rm i} [\sigma_{1A},H_{{r}}]-\frac{1}{2}(\gamma_{1A}+\gamma_{2A})\sigma_{1A},\\
    \dot{\sigma}_{2A}&=-{\rm i} [\sigma_{2A},H_{{r}}]-\frac{1}{2}(\gamma_{1A}+\gamma_{2A})\sigma_{2A},\\
    \dot{\sigma}_{1B}&=-{\rm i} [\sigma_{1B},H_{{r}}]-\frac{1}{2}(\gamma_{1B}+\gamma_{2B})\sigma_{1B},\\
    \dot{\sigma}_{2B}&=-{\rm i} [\sigma_{2B},H_{{r}}]-\frac{1}{2}(\gamma_{1B}+\gamma_{2B})\sigma_{2B},\\
    \dot{\sigma}_{11}&=-{\rm i} [\sigma_{11},H_{{r}}]+\gamma_{1A}\sigma_{AA}+\gamma_{1B}\sigma_{BB},\\
    \dot{\sigma}_{22}&=-{\rm i} [\sigma_{22},H_{{r}}]+\gamma_{2A}\sigma_{AA}+\gamma_{2B}\sigma_{BB},\\
    \dot{\sigma}_{AA}&=-{\rm i} [\sigma_{AA},H_{{r}}]-(\gamma_{1A}+\gamma_{2A})\sigma_{AA},\\
    \dot{\sigma}_{11}+&\dot{\sigma}_{22}+\dot{\sigma}_{AA}+\dot{\sigma}_{BB}=0.
\end{align}
For compact notation, we remove the symbol $\langle\cdot\rangle$. Due to the assumption of a single-sided cavity, it holds $\kappa_l=\kappa$.
The Heisenberg-Langevin equations are now a nonlinear and time-varying dynamical system. They read
\begin{align}
    &\dot{a}=-{\rm i}(g_{1A}\sigma_{1A}+e^{{\rm i}\omega_s t}g_{2A}\sigma_{2A}+e^{-{\rm i}\omega_s t}g_{1B}\sigma_{1B}+g_{2B}\sigma_{2B})-\kappa a+\sqrt{2\kappa}a_{\rm in}\label{eq:adot},\\
    &\dot{\sigma}_{1A}=-{\rm i}(\delta\sigma_{1A}+e^{-{\rm i}\omega_s t}g_{2A}^*\sigma_{1A}\sigma_{2A}^\dagger a-e^{{\rm i}\omega_s t}g_{1B}^* \sigma_{1B}^\dagger\sigma_{1A} a+g_{1A}^*a(\sigma_{11}-\sigma_{AA}))-\frac{1}{2}(\gamma_{1A}+\gamma_{2A})\sigma_{1A},\\
    &\dot{\sigma}_{2A}=-{\rm i}(\delta\sigma_{2A}+g_{1A}^* \sigma_{2A}\sigma_{1A}^\dagger a-g_{2B}^*\sigma_{2B}^\dagger\sigma_{2A} a+e^{-{\rm i}\omega_s t}g_{2A}^*a(\sigma_{22}-\sigma_{AA}))-\frac{1}{2}(\gamma_{1A}+\gamma_{2A})\sigma_{2A},\\
    &\dot{\sigma}_{1B}=-{\rm i}(-(\omega_s-\epsilon_B+\epsilon_A-\delta)\sigma_{1B}+g_{1A}^* \sigma_{1A}^\dagger\sigma_{1B} a+g_{2B}^*\sigma_{1B}\sigma_{2B}^\dagger a+e^{{\rm i}\omega_s t}g_{1B}^* a(\sigma_{11}-\sigma_{BB}))-\frac{1}{2}(\gamma_{1B}+\gamma_{2B})\sigma_{1B},\\
    &\dot{\sigma}_{2B}=-{\rm i}(-(\omega_s-\epsilon_B+\epsilon_A-\delta)\sigma_{2B}-e^{-{\rm i}\omega_s t}g_{2A}^*\sigma_{2A}^\dagger\sigma_{2B} a+e^{{\rm i}\omega_s t }\sigma_{2B}\sigma_{1B}^\dagger g_{1B}^* a+g_{2B}^* a(\sigma_{22}-\sigma_{BB}))-\frac{1}{2}(\gamma_{1B}+\gamma_{2B})\sigma_{2B},\\
    &\dot{\sigma}_{11}=-{\rm i}(g_{1A}\sigma_{1A}a^\dagger-g_{1A}^*\sigma_{1A}^\dagger a+e^{-{\rm i}\omega_s t}g_{1B}\sigma_{1B}a^\dagger-e^{{\rm i}\omega_s t}g_{1B}^*\sigma_{1B}^\dagger a)+\gamma_{1A}\sigma_{AA}+\gamma_{1B}\sigma_{BB},\\
    &\dot{\sigma}_{22}=-{\rm i}(e^{{\rm i}\omega_s t}g_{2A}\sigma_{2A}a^\dagger-e^{-{\rm i}\omega_s t}g_{2A}^*\sigma_{2A}^\dagger a+g_{2B}\sigma_{2B}a^\dagger-g_{2B}^*\sigma_{2B}^\dagger a)+\gamma_{2A}\sigma_{AA}+\gamma_{2B}\sigma_{BB},\\
    &\dot{\sigma}_{AA}=-{\rm i}(-g_{1A}\sigma_{1A}a^\dagger+g_{1A}^*\sigma_{1A}^\dagger a-e^{{\rm i}\omega_st}g_{2A}\sigma_{2A}a^\dagger+e^{-{\rm i}\omega_s t}g_{2A}^*\sigma_{2A}^\dagger a)-(\gamma_{1A}+\gamma_{2A})\sigma_{AA},\\
    &\dot{\sigma}_{11}+\dot{\sigma}_{22}+\dot{\sigma}_{AA}+\dot{\sigma}_{BB}=0.
\end{align}
Additionally, it holds the input-output relation for the cavity
\begin{equation}
    a_{{\rm out}}+a_{{\rm in}}=\sqrt{2\kappa}a\label{eq:aout}\;.
\end{equation}
It is important to mention that $\kappa$ is a half-width at half maximum (HWHM). If it were a full-width at half maximum, the factor two would be removed.
The boundary condition for the operators depends on the initial states. For example, for the system initialized in a photonic vacuum and spin state $\ket{1}$, we can evaluate the operator $\sigma_{11}(t=0)=1$ and $\sigma_{22}(t=0)=0$. Additionally, when calculating the outcoming mode for a broadband incoming photon, we assume an incoming mode $a_{\rm in}(t)=e_0 e^{({\rm i}\omega_0-\gamma/2) t}$ with amplitude $e_0$, central frequency $\omega_0$ and bandwidth $\gamma$. 

If the rotating wave approximation (RWA) allows for neglecting cross couplings, the spin-dependent reflection is modeled using the reflection coefficients. 
\begin{align}\label{eq:refl_coeff}
    R_{1}(\omega)&=-1+\frac{2 \kappa({\rm i}(\omega-\omega_{1A})+\gamma_{\rm avg,A})}{({\rm i}(\omega-\omega_c)+\kappa)({\rm i}(\omega-\omega_{1A})+\gamma_{\rm avg,A})+\vert g_{1A}\vert^2},\\
    R_{2}(\omega)&=-1+\frac{2 \kappa ({\rm i}(\omega-\omega_{2B})+\gamma_{\rm avg,B})}{({\rm i}(\omega-\omega_c)+\kappa)({\rm i}(\omega-\omega_{2B})+\gamma_{\rm avg,B})+\vert g_{2B}\vert^2},
\end{align}
with $\gamma_{\rm avg,A}=\frac{1}{2}(\gamma_{1A}+\gamma_{2A})$ and $\gamma_{\rm avg,B}=\frac{1}{2}(\gamma_{1B}+\gamma_{2B})$. Since $\gamma_{1A}\gg \gamma_{2A}$ and $\gamma_{2B}\gg\gamma_{1B}$ it is sufficient to set $\gamma_{\rm avg,A}=\frac{1}{2}\gamma_{1A}$ and $\gamma_{\rm avg,B}=\frac{1}{2}\gamma_{2B}$. Previous studies derived the reflection spectrum of a three-level atom~\cite{reiserer_cavity-based_2015} and G4V~\cite{omlor_entanglement_2024}. It is important to point out that $\kappa$ is a half-width at half maximum. In \cite{reiserer_cavity-based_2015}, the atomic decay rate $\gamma:=\gamma_{1A}/2$ is used, meaning that $\gamma$ is a half-width at half maximum as well. Note that $\kappa$ and $\gamma$ can be defined either as the half-width at half maximum (HWHM) or the full-width at half maximum (FWHM); care should be taken to ensure consistency.

In \cite{omlor_entanglement_2024}, the assumed magnetic field is assumed to be close to the defect's symmetry axis such that the coupling strengths $g_{2A},g_{1B}$ are small enough to neglect them. In our modeling, we include the cross-talk effects and solve the system of equations above with numerical methods shown in \cite{2020SciPy-NMeth} for solving ordinary differential equations. 
\section{OPTIMIZING SPIN-PHOTON CPHASE GATES}
For the optimization of the controlled phase gate we neglect cross-talk. To derive the objective function we firstly elaborate on the reflection scheme for broadband incoming photons. We model broadband incoming photons in time-domain by
\begin{align}
    a(t)=\epsilon_0 e^{({\rm i}\omega_0-\gamma/2)t}.
\end{align}
The amplitude $\epsilon_0$ is so small such that no driving between the ground and excited state is steered. The spectrum is given by
\begin{align}
    \tilde{S}(\omega-\omega_0)=\frac{\epsilon_0}{{\rm i}(\omega-\omega_0)+\gamma/2}.
\end{align}
\begin{equation}\label{eq:deriv}
\begin{split}
    \ket{\psi_{{\rm ph}}}\ket{{1}}=&\int_{\mathbb{R}} \tilde{S}(\omega-\omega_0)\left(\alpha\ket{\omega}_E\ket{{1}}+ \beta\ket{\omega}_L\ket{{1}}\right){\rm d}\omega\\
    \overset{{\rm early\, reflection}}{\rightarrow} &\int_{\mathbb{R}} \tilde{S}(\omega-\omega_0)\left(R_{1}(\omega)  \alpha\ket{\omega}_E\ket{{1}}+\beta\ket{\omega}_L\ket{{1}}\right){\rm d}\omega\\
    \overset{\pi/2\, {\rm rotation}}{\rightarrow} &\int_{\mathbb{R}} \tilde{S}(\omega-\omega_0)\left(R_{1}(\omega) \frac{\alpha}{\sqrt{2}}\ket{\omega}_E(\ket{{1}}+\ket{{2}})+\frac{\beta}{\sqrt{2}}\ket{\omega}_L(\ket{{1}}+\ket{{2}})\right){\rm d}\omega\\
    \overset{{\rm late\, reflection}}{\rightarrow}
    &\int_{\mathbb{R}} \tilde{S}(\omega-\omega_0)\Bigg(R_{1}(\omega)\frac{\alpha}{\sqrt{2}}\ket{\omega}_E(\ket{{1}}+\ket{{2}})\\+& \frac{\beta}{\sqrt{2}}\ket{\omega}_L(R_{1}(\omega)\ket{{1}}+R_{2}(\omega)\ket{{2}})\Bigg){\rm d}\omega
    \end{split}
\end{equation}
Assuming that every photon gets detected with unity probability the $X$-measurement has the form \cite{propp}
\begin{align}
    \rho_{+}=\int_\mathbb{R} {}_{\omega}\langle +\vert\psi\rangle\langle\psi\vert +\rangle_{\omega}\,{\rm d}\omega,\\
    \rho_{-}=\int_\mathbb{R} {}_{\omega}\langle -\vert\psi\rangle\langle\psi\vert -\rangle_{\omega}\,{\rm d}\omega
\end{align}
with $\ket{+}_\omega=\frac{\ket{\omega}_E+\ket{\omega}_L}{\sqrt{2}}$.
The respective entries are
\begin{align}
    \langle 1\vert \rho_{\pm}\vert 1\rangle &=\frac{1}{4}\vert\alpha\pm\beta\rvert^2 I_1,\\
    \langle 1\vert \rho_{\pm}\vert 2\rangle &=\frac{1}{4}\alpha^*(\alpha\pm\beta)I_1+\frac{1}{4}\beta^* (\alpha\pm\beta)I_2,\\
    \langle 1\vert \rho_{\pm}\vert 2\rangle&=\langle 2\vert \rho_{\pm}\vert 1\rangle^*,\\
    \langle 2\vert \rho_{\pm}\vert 2\rangle &=\frac{1}{4}\vert\alpha\vert^2 I_1\pm\frac{1}{4}\alpha^* \beta I_2^*\pm\frac{1}{4}\alpha\beta^* I_2+\frac{1}{4}\vert\beta\vert^2 I_3
\end{align}
with the integrals
\begin{align}
    I_1&=\int_\mathbb{R} S(\omega-\omega_0)\vert R_1 (\omega)\vert^2\, {\rm d}\omega,\\
    I_2&=\int_\mathbb{R} S(\omega-\omega_0) R_1 (\omega)R_2^*(\omega)\,{\rm d}\omega,\\
    I_3&=\int_\mathbb{R} S(\omega-\omega_0)\vert R_2 (\omega)\vert^2\,{\rm d}\omega.
\end{align}
and $S(\omega)=\gamma/(2\pi e_0^2)\vert\tilde{S}(\omega)\vert^2$.
The total spin state reads
\begin{align}
    \rho=R_y(\pi/2)\rho_+ R_y^\dagger(\pi/2) +\sigma_z R_y(\pi/2) \rho_- R_y^\dagger(\pi/2)\sigma_z
\end{align}
with the $\pi/2$ rotation $R_y(\pi/2)=\frac{1}{\sqrt{2}}\begin{pmatrix}
    1 & -1 \\
    1 & 1
\end{pmatrix}$ and the Pauli matrix $\sigma_z=\begin{pmatrix}
    1 & 0\\
    0 & -1
\end{pmatrix}$. We apply the $\pi/2$ rotation for getting the state
\begin{align}
    \rho_{\rm tgt}=\ket{\psi}\bra{\psi}
\end{align}
with 
\begin{align}
    \ket{\psi}=\alpha\ket{2}+\beta\ket{1}
\end{align}
as the target state. We define $\eta={\rm tr}(\rho)$ as the success probability of spin-photon entanglement.
We optimize for cavity parameters $(\kappa,\omega_c)$ and emission central frequency $\omega_0$ such that the fidelity
\begin{align}\label{eq:fid}
    F=\frac{1}{\eta}\bra{\phi}\rho\ket{\phi}
\end{align}
with $\ket{\phi}=\frac{1}{\sqrt{2}}(\ket{1}+\ket{2})$ is maximized for $\alpha=\beta=1/\sqrt{2}$.
\section{READ-IN PROCESS}\label{app:readin}
For the read-in process we model a photon source emitting broadband photons of bandwidth $\gamma$ and assume an imperfect $\pi/2$ spin rotation produced by either microwave or optical spin control. In the following we use superscripts as matrix elements an use capital letters for indexing the photonic qubit $(I,K)$ and a lowercase letter $(m,k)$ for the spin qubit index.
To model the read-in process with Kraus-operators the reflection scheme is applied to each basis state, i.e.
\begin{align}
    \rho_{\rm ph}^{(IK)}=\ket{I}\bra{K}
\end{align}
with $I,K=E,L$.
We assume an imperfect spin $\pi/2$ rotation modeled by a map $\Lambda$, i.e.
\begin{align}
    \mathcal{D}_{\pi/2}(\ket{1}\bra{1})=\sum \Lambda^{(mk)} \ket{m}\bra{k}.
\end{align}
The reflection scheme entails the early reflection, a spin $\pi/2$ rotation and a late reflection and reads 
\begin{align}
    &\rho_0^{(IK)}:=\ket{I}\bra{K}1\rangle\bra{1}=a^I \bar{a}^K \ket{1}\bra{1}\\
    \xrightarrow[]{\text{early reflection}} &(\delta_{IE}\mathcal{D}_1(a^E)+\delta_{IL}a^L)(\delta_{KE}\bar{\mathcal{D}}_1(a^E)+\delta_{KL}\bar{a}^L)\ket{1}\bra{1}\\
    \xrightarrow[]{\pi/2\,\,\text{rotation}} &\sum_{m,k} \Lambda^{(mk)} (\delta_{IE}\mathcal{D}_1(a^E)+\delta_{IL}a^L)(\delta_{KE}\bar{\mathcal{D}}_1(a^E)+\delta_{KL}\bar{a}^L)\ket{m}\bra{k}\\
    \xrightarrow[]{\text{late reflection}} &\sum_{m,k} \Lambda^{(mk)} (\delta_{IE}\mathcal{D}_1(a^E)+\delta_{IL}\mathcal{D}_m(a^L))(\delta_{KE}\bar{\mathcal{D}}_1(a^E)+\delta_{KL}\bar{\mathcal{D}}_k(a^L))\ket{m}\bra{k}=:\rho_{\rm sp-ph}^{(IK)}
\end{align}
where $I,K=E,L$ and $\mathcal{D}_k(a)$ denotes the output mode $a_{\rm out}$ from the time evolution according to the Heisenberg-Langevin equations shown in Eqs. \eqref{eq:adot}-\eqref{eq:aout} initializing $\sigma_{kk}(0)=1$ for an input mode $a_{\rm in}$.
We measure in the $X$-basis which has the basis states $\ket{\pm}=\frac{\ket{E}\pm\ket{L}}{\sqrt{2}}$.
After the $X$-measurement the state is
\begin{align}
    \rho_\pm^{(IK)}=\langle \pm\vert\rho_{\rm sp-ph}^{(IK)}\vert\pm\rangle=\frac{1}{2}\sum_{m,k}\Lambda^{(mk)}\left(\delta_{IE}\delta_{KE}I_1\pm \delta_{IE}\delta_{KL}I_k\pm\delta_{IL}\delta_{KE}I_{2m-1}+\delta_{IL}\delta_{KL}I_{2(m-1)+k}\right)\ket{m}\bra{k}
\end{align}
with $\mathbf{I}=(I_1,I_2,I_2^*,I_3)$ and 
\begin{align}
    &I_{1}=\mathcal{N}\int_0^T \vert \mathcal{D}_1 (a_{\rm in}(t))\vert^2 \,{\rm d}t,\\
    &I_{2}=\mathcal{N}\int_0^T \mathcal{D}_1 (a_{\rm in}(t))\mathcal{D}_2^* (a_{\rm in}(t))\,{\rm d}t,\\
    &I_{3}=\mathcal{N}\int_0^T \vert \mathcal{D}_2 (a_{\rm in}(t))\vert^2 \,{\rm d}t
\end{align}
where $\mathcal{N}=\gamma/e_0^2$ and $T$ sufficiently large. We compute Kraus-operators for the measurement in $\ket{+}$ and $\ket{-}$ separately using the Choi-Jamiolkowski-Isomorphism \cite{Bongioanni2010,Bhandari2016,johnston}.
\section{READ-OUT PROCESS}\label{app:readout}
To model the read-out process we calculate Kraus-operators which map the spin state to the photonic state up to a rotation. To achieve that aim we assume that the auxiliary photon source emits a photon with the fidelity $F_{\rm aux}$. The ideal state is $\ket{\psi_{\rm ph}}=1/\sqrt{2}(\ket{E}+\ket{L})$. The depolarized photonic qubit is \cite{collins_depolarizing-channel_2015}
\begin{align}
    \rho_{\rm ph}=(1-\epsilon)\ket{\psi_{\rm ph}}\bra{\psi_{\rm ph}}+\epsilon I/2,\quad\epsilon=2(1-F_{\rm aux}).
\end{align}
We perform the reflection scheme on the spin-photon state $\rho_0=\rho_{\rm ph}\otimes \ket{m}\bra{n}$ for $m,n=1,2$ and perform a $Z$-measurement on the spin qubit. We assume that the $\pi/2$ rotated spin state of $\ket{m}\bra{n}$ is given by 
$\mathcal{D}_{\pi/2}(\ket{m}\bra{n})=\sum_{p,q=1}^2 \Lambda_{mn}^{(pq)}\ket{p}\bra{q}$. The reflection scheme reads
\begin{align}
    &\rho_0=\sum_{I,K} \rho_{\rm ph}^{(IK)} a^I \bar{a}^K \ket{m}\bra{n}\\
    &\xrightarrow[]{\text{early reflection}} \sum_{I,K} \rho_{\rm ph}^{(IK)} (\delta_{IE}\mathcal{D}_m (a^E)+\delta_{IL}a^L)(\delta_{KE}\bar{D}_n (a^K)+\delta_{KL}\bar{a}^L)\ket{m}\bra{n}\\
    &\xrightarrow[]{\pi/2\,\,\text{rotation}} \sum_{I,K}\sum_{m,n} \rho_{\rm ph}^{(IK)} (\delta_{IE}\mathcal{D}_m (a^E)+\delta_{IL}a^L)(\delta_{KE}\bar{D}_n (a^K)+\delta_{KL}\bar{a}^L)\Lambda_{mn}^{pq}\ket{p}\bra{q}\\
    &\xrightarrow[]{\text{late reflection}} \sum_{I,K}\sum_{m,n} \rho_{\rm ph}^{(IK)} (\delta_{IE}\mathcal{D}_m (a^E)+\delta_{IL}\mathcal{D}_p (a^L))(\delta_{KE}\bar{D}_n (a^K)+\delta_{KL}\bar{\mathcal{D}_q}(a^L)\Lambda_{mn}^{pq}\ket{p}\bra{q}=:\rho_{\rm sp-ph}
\end{align}
We perform a $Z$-measurement of the spin qubit. It is
\begin{align}
    &\rho_1=\langle 1\vert \rho_{\rm sp-ph}\vert 1\rangle=\sum_{I,K} \rho_{\rm ph}^{(IK)} (\delta_{IE}\mathcal{D}_m (a^E)+\delta_{IL}\mathcal{D}_1 (a^L))(\delta_{KE}\bar{D}_n (a^K)+\delta_{KL}\bar{\mathcal{D}_1}(a^L)\Lambda_{mn}^{(11)},\\
    &\rho_2=\langle 2\vert \rho_{\rm sp-ph}\vert 2\rangle=\sum_{I,K} \rho_{\rm ph}^{(IK)} (\delta_{IE}\mathcal{D}_m (a^E)+\delta_{IL}\mathcal{D}_2 (a^L))(\delta_{KE}\bar{D}_n (a^K)+\delta_{KL}\bar{\mathcal{D}_2}(a^L)\Lambda_{mn}^{(22)}.
\end{align}
To evaluate the components of the photonic qubits $\rho_1$ and $\rho_2$ in the time-bin basis $\{\ket{E},\ket{L}\}$ we evaluate the state components, i.e.
\begin{align}
    &\langle E\vert\rho_i\vert E\rangle=\mathcal{N}\int_0^T \Lambda_{mn}^{(ii)}\rho_{\rm ph}^{(EE)}\mathcal{D}_m(a_{\rm in})\bar{\mathcal{D}}_n (a_{\rm in})\,{\rm d}t=\mathcal{N}\Lambda_{mn}^{(ii)}\rho_{\rm ph}^{(EE)} I_{2(m-1)+n},\\
    &\langle E\vert\rho_i\vert L\rangle=\mathcal{N}\int_0^T \Lambda_{mn}^{(ii)}\rho_{\rm ph}^{(EL)}\mathcal{D}_m(a_{\rm in})\bar{\mathcal{D}}_i (a_{\rm in})\,{\rm d}t=\mathcal{N}\Lambda_{mn}^{(ii)}\rho_{\rm ph}^{(EL)} I_{2(m-1)+i},\\
    &\langle L\vert\rho_i\vert E\rangle=\mathcal{N}\int_0^T \Lambda_{mn}^{(ii)}\rho_{\rm ph}^{(LE)}\mathcal{D}_i(a_{\rm in})\bar{\mathcal{D}}_n (a_{\rm in})\,{\rm d}t=\mathcal{N}\Lambda_{mn}^{(ii)}\rho_{\rm ph}^{(LE)} I_{2(i-1)+n},\\
    &\langle L\vert\rho_i\vert L\rangle=\mathcal{N}\int_0^T \Lambda_{mn}^{(ii)}\rho_{\rm ph}^{(LL)}\vert\mathcal{D}_i(a_{\rm in})\vert^2\,{\rm d}t=\mathcal{N}\Lambda_{mn}^{(ii)}\rho_{\rm ph}^{LL} I_{3(i-1)+1}
\end{align}
for $i=1,2$ where $\textbf{I}$ is shown in the Appendix \ref{app:readin}.
We compute Kraus-operators for the measurement in $\ket{1}$ and $\ket{2}$ separately using the Choi-Jamiolkowski-Isomorphism \cite{Bongioanni2010,Bhandari2016,johnston}.
\section{QUANTUM MEMORY PERFORMANCE}\label{app:performance}
To estimate the quantum memory performance we evaluate the average trace and infidelity for the read-in and read-out process of a photonic qubit.
We evaluate the following steps:
\begin{enumerate}
    \item Compute Kraus-operators $\{K_i^{+},K_i^{-}\}$ for the read-in process of any photonic qubit
    \item Compute Kraus-operators $\{K_i^{1},K_i^{2}\}$ for the read-out process of the spin qubit
    \item Consider the photonic qubit
    \begin{align}
    \ket{\psi_{\rm ph}(\theta,\phi)}=\cos(\theta/2)\ket{E}+e^{{\rm i}\phi}\sin(\theta/2)\ket{L},\quad \theta\in [0,\pi],\phi\in [0,2\pi)
    \end{align}
    with the respective density matrix $\rho_{\rm ph}(\theta,\phi)=\ket{\psi_{\rm ph}(\theta,\phi)}\bra{\psi_{\rm ph}(\theta,\phi)}$. We evaluate the spin states after the read-in process, i.e.
    \begin{align}
        \rho_+(\theta,\phi)=\sum_{i=1}^4 K_i^+\rho_{\rm ph}(\theta,\phi) (K_i^+)^\dagger,\\
        \rho_-(\theta,\phi)=\sum_{i=1}^4 K_i^-\rho_{\rm ph}(\theta,\phi) (K_i^-)^\dagger.
    \end{align}
    Using the auxiliary photon source we use the reflection scheme to entangle the spin states $\rho_+$ and $\rho_-$ with the auxiliary photon. After the read-out process the photonic states are
    \begin{align}
        \rho_{+,1}(\theta,\phi)=\sum_{i=1}^4 K_i^1\rho_{\rm +}(\theta,\phi) (K_i^1)^\dagger,\\
        \rho_{+,2}(\theta,\phi)=\sum_{i=1}^4 K_i^2\rho_{\rm +}(\theta,\phi) (K_i^2)^\dagger,\\
        \rho_{-,1}(\theta,\phi)=\sum_{i=1}^4 K_i^1\rho_{\rm -}(\theta,\phi) (K_i^1)^\dagger,\\
        \rho_{-,2}(\theta,\phi)=\sum_{i=1}^4 K_i^2\rho_{\rm -}(\theta,\phi) (K_i^2)^\dagger.
    \end{align}
    The total photonic states are
    \begin{align}
        \rho_{\rm ph,+}(\theta,\phi)=\rho_{+,1}(\theta,\phi)+\sigma_z\sigma_x\rho_{+,2}(\theta,\phi)\sigma_x\sigma_z,\\
        \rho_{\rm ph,-}(\theta,\phi)=\rho_{-,1}(\theta,\phi)+\sigma_z\sigma_x\rho_{-,2}(\theta,\phi)\sigma_x\sigma_z
    \end{align}
    with $\sigma_x=\begin{pmatrix}
        0 & 1 \\
        1 & 0
    \end{pmatrix}$ and $\sigma_z=\begin{pmatrix}
        1 & 0 \\
        0 & -1
    \end{pmatrix}$.
    From these states the total photonic state after read-out is given by
    \begin{align}
        \tilde{\rho}_{\rm ph}(\theta,\phi)=\rho_{\rm ph,+}(\theta,\phi)+\sigma_z\rho_{\rm ph,-}(\theta,\phi)\sigma_z
    \end{align}
    The trace of the photon after the joint read-in and read-out process is subsequently given by ${\rm tr}(\tilde{\rho}_{\rm ph}(\theta,\phi))$ and the fidelity is $F(\theta,\phi)=\frac{1}{\eta(\theta,\phi)}\langle\psi_{\rm ph}(\theta,\phi)\vert\tilde{\rho}_{\rm ph}(\theta,\phi)\vert\psi_{\rm ph}(\theta,\phi)\rangle$.
    \item Compute the Bloch average for the fidelity and trace
    \begin{align}
        \langle F\rangle:=\frac{1}{4\pi}\int_0^\pi \int_0^{2\pi} F(\theta,\phi)\sin(\theta)\,{\rm d}\theta\,{\rm d}\phi,\quad \langle {\rm tr}(\rho_{\rm ph})\rangle:=\frac{1}{4\pi}\int_0^\pi \int_0^{2\pi} {\rm tr}(\tilde{\rho}_{\rm ph}(\theta,\phi))\sin(\theta)\,{\rm d}\theta\,{\rm d}\phi.
    \end{align}

\end{enumerate}
\section{Supporting Data}
    \begin{table}[h]
    \centering
    \caption{Microwave powers and total processing times for qubit storage and retrieval $(P^{\rm MW}/\mu\text{W},T_1/{\rm ns})$ assuming a fiber length $L=100$ m, a detector's dead time $T_m=100$ ps, the refractive index of diamond $n=2.417$, the bandwidth of the photon for read-in and read-out $\gamma_{\rm in,out}=0.1$ GHz, for $\theta_{\rm dc}=\pi/2$, $\theta_{\rm ac}=0$ with the field strength $B_{\rm ac}=1.0$ $(3.7)$ mT for the SiV$^{-}$(SnV$^{-}$)as quantum memory}
    \begin{tabular}{cccc}
       Center & $B_{\rm dc\perp}$ (T)  & $P^{\rm MW}$ $(\mu\text{W})$ & $T_1$ (ns)\\
       \hline
        SiV$^{-}$& $0.3$ & $359$ & $509$\\
        SiV$^{-}$& $0.4$ & $202$ & $509$\\
        SnV$^{-}$& $0.3$ & $876$ & $531$\\
        SnV$^{-}$& $0.4$ & $493$ & $531$
    \end{tabular}
    \label{tab:summary}
\end{table}
\begin{table}[h]
    \centering
    \caption{Microwave powers and total processing times for qubit storage and retrieval $(P^{\rm MW}/\mu\text{W},T_1/{\rm ns})$ assuming a fiber length $L=100$ m, a detector's dead time $T_m=100$ ps, the refractive index of diamond $n=2.417$, the bandwidth of the photon for read-in and read-out $\gamma_{\rm in,out}=0.1$ GHz, for $\theta_{\rm dc}=0$, $\theta_{\rm ac}=\pi/2$ with the field strength $B_{\rm dc}=0.3$ T and $B_{\rm ac}=1.0$ $(3.7)$ mT for the SiV$^{-}$ (SnV$^{-}$) as quantum memory}
    \begin{tabular}{ccc}
       &$P^{\rm MW}$ $(\mu\text{W})$ & $T_1$ $({\rm ns})$\\
       \hline
        SiV$^{-}$& $274$ & $539$\\
        SnV$^{-}$& $16$ & $2802$
    \end{tabular}
    \label{tab:summary}
\end{table}
\begin{table}[h]
    \centering
    \caption{Laser power saturation ratio and total processing times for qubit storage and retrieval $(s_1,s_2,T_1/{\rm ns})$ assuming a fiber length $L=100$ m, a detector's dead time $T_m=100$ ps, the refractive index of diamond $n=2.417$, the bandwidth of the photon for read-in and read-out $\gamma_{\rm in,out}=0.1$ GHz.}
    \begin{tabular}{cccc}
       $B_{\rm dc}$ (T) & $0.3$ & $1.0$ & $3.0$\\
       \hline
        $s_1$ & $0.430$ & $0.744$ & $0.093$\\
        $s_2$ & $0.042$ & $0.066$ & $0.004$\\
        $T_1$ (ns) & $501.29$ & $500.55$ & $503.00$
    \end{tabular}
    \label{tab:summary}
\end{table}
\end{appendix}
\end{document}